\documentclass{raa}            % referee version: for submission
\usepackage{graphicx,times}             %for PS/EPS graphics inclusion, new
\usepackage{amssymb}

\begin{document}

   \title{Galactic Disk Bulk Motions as Revealed by the LSS-GAC DR2
%\footnotetext{$*$ Supported by the National Natural Science Foundation of China.}
}
%   \subtitle{I. Place Your Subtitle Here}

   \volnopage{Vol.0 (200x) No.0, 000--000}      %%preserved for Editor. DOn't remove!
   \setcounter{page}{1}          %%starting page, preserved for Editor. DOn't remove!

   \author{Ning-Chen Sun
      \inst{1}
   \and Xiao-Wei Liu
         \inst{1,2}
   \and Yang Huang
      \inst{1}
   \and Hai-Bo Yuan\footnotemark[1] \footnotetext{\footnotemark[1] LAMOST Fellow}
      \inst{2}
   \and Mao-Sheng Xiang
      \inst{1}
   \and Hua-Wei Zhang
      \inst{1}
   \and Bing-Qiu Chen\footnotemark[1]
      \inst{1}
   \and Juan-Juan Ren\footnotemark[1]
      \inst{1}
   \and Chun Wang
      \inst{1}
   \and Yong Zhang
      \inst{3}
   \and Yong-Hui Hou
      \inst{3}
   \and Yue-Fei Wang
      \inst{3}
    \and Ming Yang
      \inst{4}
    }
   \institute{Department of Astronomy, Peking University, Beijing 100871, People's Republic of China;
   {\it sunnc.astro@pku.edu.cn (NCS); x.liu@pku.edu.cn (XWL); yanghuang@pku.edu.cn (YH)}\\
        \and
             Kavli Institute for Astronomy and Astrophysics, Peking University, Beijing 100871, People's Republic of China\\
        \and
             Nanjing Institute of Astronomical Optics \& Technology, National Astronomical Observatories, Chinese Academy of Sciences, Nanjing 210042, China\\
        \and
             Key Laboratory of Optical Astronomy, National Astronomical Observatories, Chinese Academy of Sciences, Beijing 100012, China\\
   }

%   \date{Received~~2009 month day; accepted~~2009~~month day}

\abstract{We report a detailed investigation of the bulk motions of the nearby Galactic stellar disk, based on three samples selected from the LSS-GAC DR2: a global sample containing 0.57 million FGK dwarfs out to $\sim$~2~kpc, a local subset of the global sample consisting $\sim$~5,400 stars within 150~pc, and an anti-center sample containing $\sim$~4,400 AFGK dwarfs and red clump stars within windows of a few degree wide centered on the Galactic anti-center. The global sample is used to construct a three-dimensional map of bulk motions of the Galactic disk from the solar vicinity out to $\sim$~2~kpc with a spatial resolution of $\sim$~250~pc. Typical values of the radial and vertical components of bulk motion range from $-$15~km~s$^{-1}$ to 15~km~s$^{-1}$, while the lag behind the circular speed dominates the azimuthal component by up to $\sim$~15~km~s$^{-1}$. The map reveals spatially coherent, kpc-scale stellar flows in the disk, with typical velocities of a few tens~km~s$^{-1}$. Bending- and breathing-mode perturbations are clearly visible, and vary smoothly across the disk plane. Our data also reveal higher-order perturbations, such as breaks and ripples, in the profiles of vertical motion versus height. From the local sample, we find that stars of different populations exhibit very different patterns of bulk motion. Finally, the anti-center sample reveals a number of peaks in stellar number density in the line-of-sight velocity versus distance distribution, with the nearer ones apparently related to the known moving groups. The ``velocity bifurcation" reported by Liu et al. (2012) at Galactocentric radii 10--11~kpc is confirmed. However, just beyond this distance, our data also reveal a new triple-peaked structure.
}

   \authorrunning{N.-C. Sun, X.-W. Liu \& Y. Huang et al.}            %author_head in even pages
   \titlerunning{Galactic Disk Bulk Motions}  % title_head in odd pages

   \maketitle
%% The author head (on even pages) and the title head (on odd pages) will be
%% automatically extracted from \author{} and \title{}. Whenever the title is too long,
%% you will be asked to supply a shorter one by inserting either \authorrunning{} or
%% \titlerunning{} before \maketitle. Anyway, you can specify your own heads.
%%
%%
%% Note: In the following text body of your manuscript, please note several differences from
%%       other major journals:
%% (1) \subsection{Please Capitalize the First Letter of Each Notional Word in Subsection Title}
%% (2) Please Capitalize the First Letter of Each Notional Word in all tables' captions

%
%________________________________________________ sections below
%
\section{Introduction}           %% first-level sections will be auto-capitalized
\label{sec_intro}

The stellar velocity distributions encode the assemblage history and structural information of the Milky Way. The Milky Way is often approximately considered to be steady and axisymmetric. In such a model, disk stars are expected to have quite smooth and single-peaked velocity distributions, without significant bulk motions in either the radial or vertical directions (see e.g. Schwarzschild \& Villiger 1907; Sch{\"o}nrich \& Binney 2012).
%with no bulk motions

This simple picture is, however, challenged by the recent detections of significant bulk motions in the nearby stellar disk. Using $\sim$~1$\times$10$^4$ stars from the Sloan Extension for Galactic Understanding and Exploration survey (SEGUE; Yanny et al.~2009), Widrow et al.~(2012) determine the vertical bulk motions as a function of height from the disk mid-plane. Their results reveal wave-like perturbations and some differences in bulk motion between blue and red stars. Williams et al.~(2013) report the bulk motions of $\sim$~7$\times$10$^4$ red clump stars selected from the Radial Velocity Experiment (RAVE; Steinmetz et al.~2006). They find a north-south asymmetry both in the radial and vertical bulk motions; and curiously, the vertical bulk motions exhibit some compression and rarefaction patterns across the disk. North-south asymmetry and compression patterns are also observed by Carlin et al.~(2013) with $\sim$~4$\times$10$^5$ F-type stars selected from the LAMOST Galactic spectroscopic surveys (Cui et al.~2012; Zhao et al.~2012) and RAVE. Bovy et al.~(2015) report fluctuations in differences between the observed and predicted stellar velocity distributions on scales ranging from 25~pc to 10~kpc, using data from the Apache Point Observatory Galactic Evolution Experiment (APOGEE; Eisenstein et al.~2011), RAVE and the Geneva-Copenhagen Survey (GCS; Nordstr{\"o}m et al.~2004). They attribute the fluctuations to the effects of the central bar. Besides, Liu et al.~(2012) reveal a feature of ``velocity bifurcation" at Galactocentric radii 10--11~kpc with $\sim$~700 red clump stars in the Galactic anti-center direction, and interpret it as a resonance feature of the central bar. Theoretically, Siebert et al.~(2011), Debattista~(2014) and Faure et al.~(2014) show that bulk motions can be excited by spiral arms, and Widrow et al.~(2014) investigate the dynamical effects of a passing satellite or dark matter subhalo.
%report a feature of

By June 2014, the LAMOST Spectroscopic Survey of Galactic Anti-Center (LSS-GAC; Liu et al.~2014; Yuan et al.~2015) has obtained spectra with S/N(4650\AA)~$\geq$~10 for over one million stars. The second data release of value-added catalogs (DR2) provides their line-of-sight radial velocities, stellar atmospheric parameters, three-dimensional extinctions, distances and proper motions. In this paper, we report an investigation of bulk motions based on three samples selected from the LSS-GAC DR2: A global, a local and an anti-center sample. The global sample, containing $\sim$~0.57 million FGK dwarfs, is used to obtain a three-dimensional map of bulk motions of the Galactic disk from the solar neighborhood out to $\sim$~2~kpc with a resolution of $\sim$~250~pc. The second sample, a subset of the global one, contains $\sim$~5,400 stars within 150~pc; we use it to investigate the bulk motions of different stellar populations in the solar neighborhood. Finally, we investigate whether the ``velocity bifurcation" reported by Liu et al.~(2012) is a localized structure via an analysis of the line-of-sight radial velocities of the anti-center sample, consisting $\sim$~4,400 AFGK dwarfs and red clump giants, tracing a contiguous distance range of 4~kpc in the Galactic anti-center direction. The paper is organized as follows. After a description of the data in Section~\ref{sec_data}, we present results based on the three samples in Sections~\ref{sec_global}--\ref{sec_popgac} in turn. We then close with a brief summary in Section~\ref{sec_sum}.

%% Authors can give a citation as 'Michel et al. 1992'.
%% You may also use \cite, \citep and \citet for citation, and use Table~1 or Figure~1
%% and so forth. Using \ref and \label for cross-references of Tables/Figures
%% is a good way in adjusting/adding/removing text, tables or figures.

\section{Data}
\label{sec_data}

\subsection{LSS-GAC}
\label{sec_lssgac}

Data used in this work are from the LSS-GAC DR2 (Xiang et al.~2015b, in preparation). LSS-GAC -- the LAMOST Spectroscopic Survey of the Galactic Anti-Center (Liu et al.~2014, Yuan et al.~2015) surveys a 3400~deg$^2$ contiguous area centered on the Galactic anti-center (150$^\circ$~$<~l~<$~210$^\circ$, $-$30$^\circ$~$<$~$b$~$<$~30$^\circ$) and aims to obtain optical ($\lambda\lambda$~3700~$-$~9000), low-resolution ($R$~$\sim$~1,800) spectra for a statistically complete sample\footnote{In the sense that the sample stars are selected uniformly on celestial sphere to given limiting magnitudes and randomly in color-magnitude diagrams, such that with due considerations of the various well-defined selection effects, the spectroscopically targeted sample stars can be used to recover the underlying photometric populations (cf. Yuan et al. 2015 for detail).} containing $\sim$ 3 million stars of all colors and of magnitudes ranging from $r$~=~14.0 to 17.8~mag (down to 18.5~mag for a limited number of fields). Over 1.5 million stars of $-$10$^\circ$~$<$~$Dec$~$<$~60$^\circ$ and brighter than $r$~=~14~mag, are also selected with a similar selection algorithm and observed under bright lunar conditions. LSS-GAC also contains a sub-program, targeting various types of objects in the M31-M33 region. By June, 2014, LSS-GAC has obtained over two million spectra with a signal-to-noise ratio at 4650\,{\AA}, S/N(4650\AA)~$\geq$~10.

Line-of-sight radial velocities, $V_r$, and stellar atmospheric parameters (effective temperature $T_{\rm eff}$, surface gravity log~$g$ and metallicity [Fe/H]) are estimated with the LAMOST Stellar Parameter Pipeline at Peking University (LSP3; Xiang et al.~2015a). For S/N(4650\AA)~$\geq$~10, LSP3 has achieved accuracies better than 5.0~km~s$^{-1}$, 150~K, 0.25~dex and 0.15~dex for $V_r$, $T_{\rm eff}$, log~$g$ and [Fe/H], respectively. Values of the interstellar extinction, with typical errors of $\sim$~0.04~mag, are obtained for individual stars with the standard pairing technique (Stecher~1965, Massa et al.~1983, Yuan et al.~2013). Distances have been derived based on an empirical relation between absolute magnitudes ($g_0$, $r_0$, $J_0$, $H_0$, $K_{s0}$) and stellar atmospheric parameters ($T_{\rm eff}$, log~$g$, [Fe/H]). Typical distance uncertainties are 10--15\% for dwarfs and 20--30\% for giants.

%Typical distance accuracies are 10--15\% for dwarfs or 20--30\% for giants.

Proper motions, $\mu_\alpha$cos$\delta$ and $\mu_\delta$, are from the UCAC4 (Zacharias et al.~2013) and PPMXL (Roeser et al.~2010) catalogs. UCAC4 is an all-sky, astrometric catalog containing over 113 million objects complete to $R$~$\sim$~16~mag. More than 105 million stars in UCAC4 have proper motions, with typical random errors of 4~mas~yr$^{-1}$. Compiled based on USNO-B1.0, 2MASS and a previous version of PPMX, PPMXL contains over 900 million sources down to $V$~=~20~mag. Almost all of them have proper motions, with typical random errors ranging from 4 to more than 10~mas~yr$^{-1}$. The position-dependent systematic errors of proper motions of UCAC4 and PPMXL have been corrected for by Huang et al.~(2015) and Carlin et al.~(2013), respectively, using $\sim$~1,700 or $\sim$~10$^5$ quasars as anchors. After the corrections, however, some unaccounted for color- and magnitude-dependent systematic errors remain, which are further discussed in Section~\ref{sec_err}.

\subsection{Coordinate Systems and Galactic Parameters}
\label{sec_coord}

We use two sets of coordinate systems in this paper: (a) A right-handed Cartesian coordinate system ($X$, $Y$, $Z$) centered on the Sun, with $X$ increasing towards the Galactic center, $Y$ in the direction of Galactic rotation and $Z$ the height from the disk mid-plane, positive towards the north Galactic pole; and (b) A Galactocentric cylindrical coordinate system ($R$, $\Phi$, $Z$), with $R$ the Galactocentric distance, $\Phi$ increasing in the direction of Galactic rotation and $Z$ the same as that in the Cartesian system. We assume that the Sun has a Calactocentric distance $R_\odot$~=~8~kpc. This places the Galactic center at ($X_{\rm GC}$, $Y_{\rm GC}$, $Z_{\rm GC}$) = (8, 0, 0)~kpc and the Sun at ($R_\odot$, $\Phi_\odot$, $Z_\odot$) = (8~kpc, 0$^\circ$, 0~kpc). The three velocity components are denoted as ($U$, $V$, $W$) in the Cartesian system and ($V_R$, $V_\Phi$, $V_Z$) in the cylindrical system. We use a flat rotation curve of $V_c$~=~220~km~s$^{-1}$, and solar motions ($U_\odot$, $V_\odot$, $W_\odot$) = (7.01, 10.13, 4.95)~km~s$^{-1}$ (Huang et al.~2015) relative to the Local Standard of Rest (LSR).

\subsection{Samples}
\label{sec_sample}

\subsubsection{The global sample}
\label{sec_glsam}

We use a global sample of FGK dwarfs to construct a three-dimensional map of bulk motions of the Galactic disk from the solar vicinity out to $\sim$~2~kpc. The stars are selected from the LSS-GAC DR2 by applying effective temperature and surface gravity cuts, 4,200 $<$ $T_{\rm eff}$ $<$ 6,800~K and 3.8 $<$ log~$g$ $<$ 5.0~dex. Stars without UCAC4 or PPMXL proper motions are excluded and duplicate ones of lower spectral SNRs removed. This leads to a sample of $\sim$ 0.70 million FGK dwarfs.

We derive the three-dimensional positions, ($X$, $Y$, $Z$) and ($R$, $\Phi$, $Z$), and three-dimensional velocities, ($U$, $V$, $W$) and ($V_R$, $V_\Phi$, $V_Z$), of individual stars from their celestial coordinates, ($l$, $b$), distances, $d$, line-of-sight radial velocities, $V_r$, and proper motions, $\mu_{\alpha}$cos$\delta$ and $\mu_\delta$. For proper motions, both UCAC4 and PPMXL catalogs are used, yielding two independent sets of velocities for each star. We use a Monte Carlo method of error propagation to estimate the random errors of positions and velocities. This is done by randomly sampling the distance moduli, radial velocities and proper motions, assuming Gaussian random error distributions. For each realization of a given star, we calculate the position and velocity. This is repeated 1000 times to obtain their distributions. The standard deviations of the distributions are adopted as random errors of the position and velocity of the star in concern.

\begin{figure}
\centering
\includegraphics[scale=0.5]{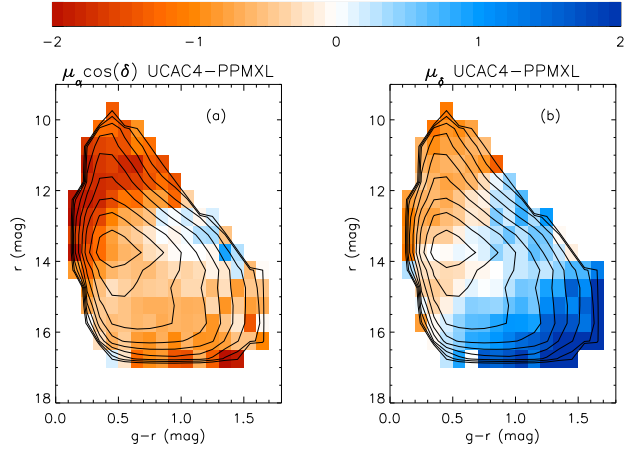}
\caption{Color-magnitude diagram of the global sample (contours), and the median differences of UCAC4 and PPMXL proper motions ($\mu_\alpha\cos(\delta)$ on the left and $\mu_\delta$ on the right) of the member stars falling in the individual color-magnitude bins (colorscale). The bins have a width of 0.1~mag in $g-r$ and 0.5~mag in $r$. The contours increase from 100 to $\sim$~1$\times$10$^4$ stars per bin with equal logarithmic steps. The colorbar is in units of mas~yr$^{-1}$.}
\label{figsys}
\end{figure}

\begin{figure*}
\centering
\includegraphics[scale=0.8]{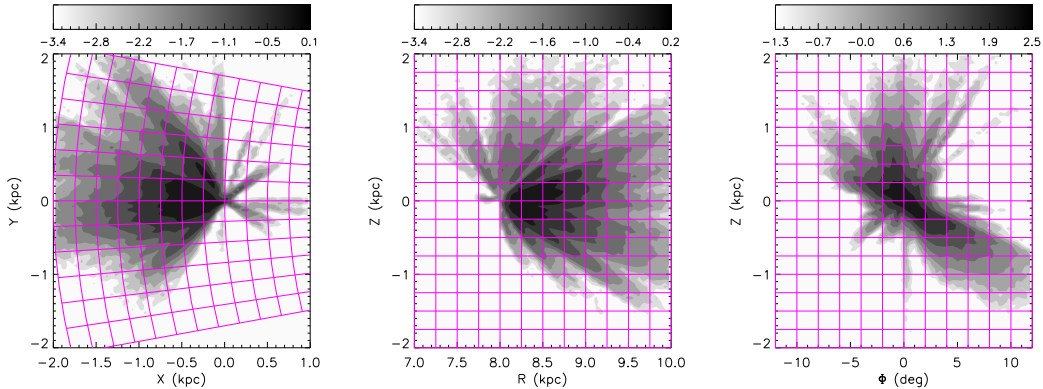}
\caption{Spatial distributions of the global sample. Greyscale shows the logarithmic surface density of stars in units of pc$^{-2}$ (left and middle panels) or in pc$^{-1}$~deg$^{-1}$ (right panel). With spacings ($\Delta R$, $\Delta \Phi$, $\Delta Z$) = (250~pc, 2$^\circ$, 250~pc), the magenta grids show how the sample is divided three-dimensionally when we construct the bulk motion maps.}
\label{figmap}
\end{figure*}

We then remove outliers and stars with large random errors by requiring: $-$200 $<$ $V_R$ $<$ 200~km~s$^{-1}$, 0 $<$ $V_\Phi$ $<$ 400~km~s$^{-1}$, $-$200 $<$ $V_Z$ $<$ 200~km~s$^{-1}$ and random errors in $V_R$, $V_\Phi$ and $V_Z$ no larger than 50~km~s$^{-1}$. Velocities obtained with the UCAC4 and PPMXL proper motions should both meet these criteria to retain a star in the sample. This leaves $\sim$~0.57 million FGK dwarfs in the final sample. The contours in Fig.~\ref{figsys} display the sample in the ($g-r$, $r$) color-magnitude diagram. The sample covers a range of 10--17~mag in $r$-band magnitude and 0.1--1.7~mag in $g - r$ color. The spatial distributions of the stars can be found in Fig.~\ref{figmap}, which shows that the stars occupy a significant volume of the Galactic disk, spanning from the solar vicinity out to $\sim$~2~kpc. The global sample is three-dimensionally sliced into small bins, with binsize ($\Delta R$, $\Delta \Phi$, $\Delta Z$) = (250~pc, 2$^\circ$, 250~pc). The entire volume of Fig.~\ref{figmap} thus contains a total of 2304 bins. Amongst them, 662 bins are found to contain stars, ranging from several to a few 10$^4$; and 252 bins have no fewer than 100 stars.

\subsubsection{The local sample}
\label{sec_lcsam}

A local sample of stars are selected by applying a distance cut, $d$~$\leq$~150~pc, to the global sample. The local sample is less affected by the potential systematic errors in proper motions due to the distance cut. The sample contains 5,443 FGK dwarfs, and is further divided into six populations according to their effective temperatures (F/G/K types) and metallicities (metal-rich and metal-poor). Table~1 (see Section~\ref{sec_poplocal}) lists the numbers of stars, along with the effective temperature and metallicity ranges of each population.

\subsubsection{The anti-center sample}
\label{sec_acsam}

The anti-center sample is consisted of four populations, the GK-, F-, A-type dwarfs and red clump (RC) giants. The dwarfs are selected from the LSS-GAC DR2 with the effective temperature and surface gravity cuts, 4,200 $<$ $T_{\rm eff}$ $<$ 6,000~K and 3.8~dex $<$ log~$g$ $<$ 5.0~dex for GK-type stars, 6,000 $<$ $T_{\rm eff}$ $<$ 6,800~K and 3.8 $<$ log~$g$ $<$ 5.0~dex for F-type stars, and 6,800 $<$ $T_{\rm eff}$ $<$ 10,000~K and 3.2 $<$ log~$g$ $<$ 5.0~dex for A-type stars. The RC giants are selected by Huang et al.~(this volume), also from the LSS-GAC DR2, based on their positions in the [Fe/H]--$T_{\rm eff}$--log $g$ and [Fe/H]--color parameter spaces. The PARSEC stellar evolution models (Bressan et al.~2012) and high-quality asteroseismology data of Kepler are used to define the location of RC giants in the parameter spaces.  The contamination and completeness of the selection algorithm are less than 5\% and greater than 90\%, respectively. RC stars thus selected have distance accuracies better than 5--10\%. From the chosen AFGK-type dwarfs and RC giants, we then select those falling in windows centered on the Galactic anti-center, $\mid$$l -$ 180$^\circ$$\mid \leq$ $\delta$, $\mid$$b$$\mid \leq$ $\delta$, where $\delta$ is set to 4$^\circ$ for GK- and F-type dwarfs, 6$^\circ$ for A-type ones and 2$^\circ$ for RC giants, a compromise between a sufficient number of stars (preferring a large $\delta$) and a good alignment between the Galactocentric and line-of-sight radial velocities (preferring a small $\delta$). The final anti-center sample contains 1,716 GK-, 1,402 F-, 359 A-type dwarfs and 958 RC giants. With increasing luminosities, these four populations of stars are able to trace the Galactic disk in the anti-center direction from the solar vicinity out to a Galactocentric distance of 12~kpc. The distance distributions of the anti-center sample can be found in Fig.~\ref{figbif} (Section~\ref{sec_popgac}).

\section{The Three-Dimensional Bulk Motion Map}
\label{sec_global}

\subsection{Characterizing Bulk Motions}
\label{sec_char}

%The global sample is three-dimensionally sliced into small bins, with binsize ($\Delta R$, $\Delta \Phi$, $\Delta Z$) = (250~pc, 2$^\circ$, 250~pc) (see Fig.~\ref{figmap}). The entire volume of Fig.~\ref{figmap} thus contains a total of 2304 bins. Amongst them, 662 bins are found to contain stars, ranging from several to a few 10$^4$; and 252 bins have no fewer than 100 stars.

\begin{figure*}
\centering
\includegraphics[scale=0.8]{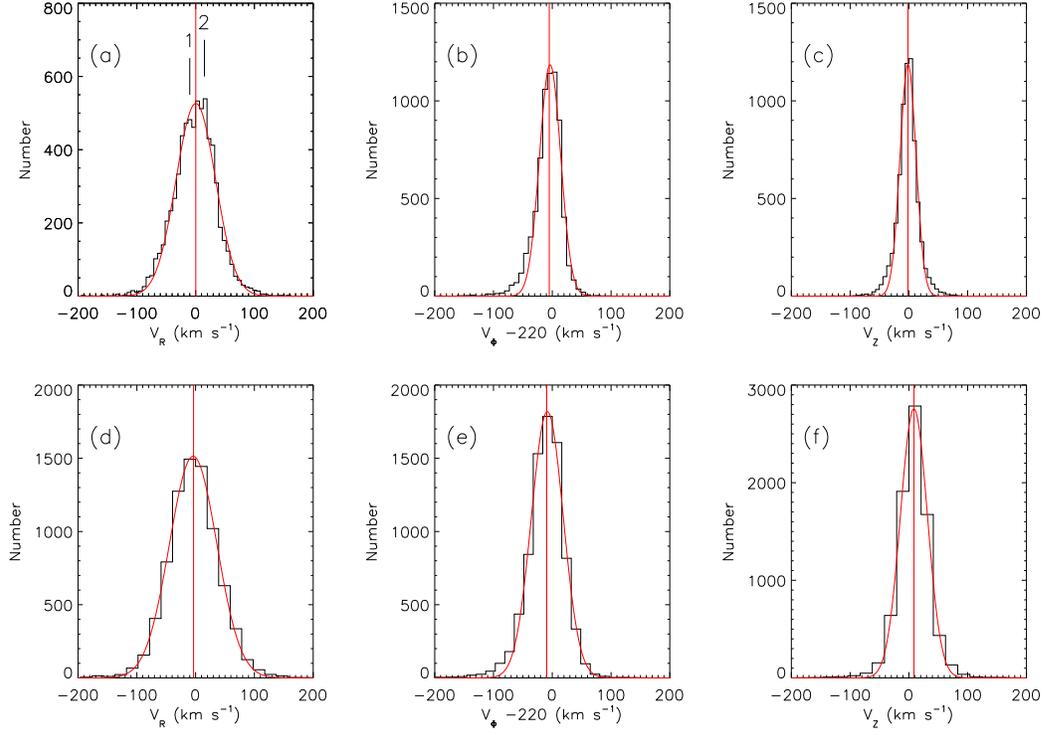}
\caption{Distributions of $V_R$ (left panels), $V_\Phi$ (middle panels) and $V_Z$ (right panels), derived using the UCAC4 proper motions in bins 8 $<$ $R$ $<$ 8.25~kpc, 0 $<$ $\Phi$ $<$ 2$^\circ$, 0 $<$ $Z$ $<$ 0.25~kpc (upper panels) and 8.25 $<$ $R$ $<$ 8.5~kpc, 4 $<$ $\Phi$ $<$ 6$^\circ$, $-$0.5 $<$ $Z$ $<$ $-$0.25~kpc (lower panels). In each panel, the histogram binsize is set to no smaller than the random errors of the 90th percentile of stars ordered by increasing error. The binsize is set in such a way to avoid artifacts introduced by the velocity errors of individual stars. The red curve is a Gaussian fit to the distribution of which the median value is marked by the red straight line. In Panel (a), Peak 1 at $-$15~km~s$^{-1}$ originates from the moving groups Sirius and UMa, and Peak 2 at 10~km~s$^{-1}$ comes from Pleiades and Coma Berenices.}
\label{figpf}
\end{figure*}

To construct the bulk motion maps, we three-dimensionally slice the global sample into small bins (see Section~\ref{sec_glsam} and Fig.~\ref{figmap}). Fig.~\ref{figpf} displays the velocity distributions of two typical bins. The first bin is a close one and thus less affected by the random errors propagated from the proper motions and distances. As Panel (a) shows, the distribution of $V_R$ in the first bin shows deviations from a simple, smooth Gaussian distribution, including two peaks at $-$15~km~s$^{-1}$ and 10~km~s$^{-1}$, respectively. The peaks are produced by the known moving groups (see e.g. Dehnen~1998; Antoja et al.~2012): the $-$15~km~s$^{-1}$ peak comes from the Sirius and UMa groups, while the one at 10~km~s$^{-1}$ from the Pleiades and Coma Berenices groups. Deviations such as those are however not so big -- a single Gaussian can still fit well the overall distribution and the median $V_R$ finds the Gaussian peak quite precisely. A visual examination of the 252 bins with no fewer than 100 stars has found clear signatures of moving groups in the $V_R$ profiles of only 4 bins, all near the Sun. Again, the presence of those moving groups hardly affect the determinations of the centroids of the distributions through the median values of $V_R$. Those moving groups, however, do not show up in $V_\Phi$ as obviously as in $V_R$ of the same bin [Panel (b)]. They are probably lost in the much narrower $V_\Phi$ distribution (The moving groups have typical widths in $V_R$ or $V_\Phi$ of $\gtrsim$~10~km~s$^{-1}$; see e.g. Fig.~3 of Dehnen~1998). Unlike $V_R$, $V_\Phi$ has long been known to skew towards lower values. This is clearly seen in the current case. Still, a single Gaussian fits the data and the median value finds the peak position well. Panel (c) shows quite a simple, single-peaked, symmetric profile of $V_Z$ of the first bin, whose peak can be easily found by Guassian fitting or by the median value despite the weak, extended wings at $\mid$$V_Z$$\mid$ $\gtrsim$ 30~km~s$^{-1}$. In the second bin of farther distance [Panels (d-f)], large random errors propagated from the proper motions and distances would have blurred the signatures of any moving groups, the skewness or the extended wings in the distributions of $V_R$, $V_\Phi$ and $V_Z$, respectively. In this case, the distribution profiles are well fitted by a Gaussian, whose peaks coincide well with the corresponding median values. Above all, as a first approximation, the bulk motions of stars in the individual bins can be characterized by the median values of the three components of velocity, denoted in this paper by $\langle V_R \rangle$, $\langle V_\Phi \rangle$ and $\langle V_Z \rangle$. To calculate those median values, we have required the bins to contain at least 100 stars, leaving 252 bins usable. The bulk motions thus derived have random errors less than $\sim$~5~km~s$^{-1}$, with typical values of $\lesssim$~2~km~s$^{-1}$ for $\langle V_R \rangle$ and $\lesssim$~3~km~s$^{-1}$ for $\langle V_\Phi \rangle$ and $\langle V_Z \rangle$.

\subsection{Systematic Errors}
\label{sec_err}

%\subsubsection{Extinctions}
%\label{sec_extin}

%Accurate corrections for the interstellar extinction are crucial for distance determinations of disk stars. Previous studies, e.g. Williams et al.~(2013) and Carlin et al.~(2013), have usually utilized the two-dimensional SFD extinction map (Schlegel et al.~1998). Although the SFD map often overestimates the true extinctions to disk stars, it introduces relatively insignificant systematic errors when the $K$-band photometry is used to estimate the distances. This is confirmed by comparing results derived for a subsample of F-type stars with $K$-band photometry. In the current work, as the distances are estimated from the magnitudes in the optical $g$, $r$ and $i$ bands, accurate three-dimensional extinction corrections are necessary. A detailed description of extinction estimation for our sample has been given in Yuan et al.~(2015).

\subsubsection{Distances}
\label{sec_errd}

The random errors of stellar distances do not affect the bulk motions systematically. Slightly evolved stars in the global sample, with 3.8~$\leq$ log~$g$ $\leq$~4.2 are found to have systematically underestimated distances by $\sim$20\% compared to those derived by isochrone fitting (Xiang M.-S. et al., private communications). However, the fraction of those evolved stars in our sample is small. Moreover, a $\sim$20\% underestimation in distances corresponds to an underestimation of the same percentage in the mean tangential velocities, which in turn contributes very minor to $\langle V_R \rangle$ and $\langle V_Z \rangle$.

\subsubsection{Line-of-Sight Radial Velocities}
\label{sec_radial}

A comparison of the line-of-sight radial velocities of common stars observed by LAMOST and APOGEE has shown that the LAMOST wavelength calibration suffers from glitches of the order of $\sim$~2~km~s$^{-1}$ on a timescale of several days (Xiang M.-S. et al., private communications). The sample stars inside each bin were observed on nights ranging from typically $\sim$~5 up to $\sim$~60 days, with typical timespans ranging from 1--3 months to 1--3 years. We expect that the effects of those glitches largely cancel out on such timescale, and thus hardly affect our estimates of the bulk motions. Unresolved binary stars can potentially affect the line-of-sight radial velocities due to the orbital velocities of the binary stars around their centers of mass. However, this effect is also unlikely to affect the determinations of the median velocities $\langle V_R \rangle$, $\langle V_\Phi \rangle$ and $\langle V_Z \rangle$ in a systematic way, under the reasonable assumption that the orbital orientations and phases of binary systems are randomly distributed.

\subsubsection{Proper Motions}
\label{sec_errpm}

The largest systematic errors affecting our results come from the proper motions, more significantly at larger distances. The colorscale in Fig.~\ref{figsys} shows the median differences of the UCAC4 and PPMXL proper motions for the global sample. It can be seen that the differences vary systematically with $g-r$ color and $r$ magnitude. Typical differences for $\mu_\alpha$cos$\delta$ range from $-$1 to 0~mas~yr$^{-1}$ except for very blue or bright stars ($g-r <$ 0.3~mag or $r <$ 12.5~mag), for which the differences amount to $-$2~mas~yr$^{-1}$. For $\mu_\delta$, the systematic differences typically range from $-$1 to 1~mas~yr$^{-1}$; but very red or faint stars ($g-r >$ 1.3~mag or $r >$ 16.0~mag) are found to have values up to 2~mas~yr$^{-1}$. Thus, even after corrected for systematics using samples of distant quasars, the proper motions from UCAC4 and PPMXL still suffer from the unaccounted for color- and magnitude-dependent systematic errors. In the current work, we have used proper motions from both catalogs and deem result robust if it is seen using both UCAC4 and PPMXL.

\subsubsection{Algorithm Biases}
\label{sec_bias}

As we shall see later, $\langle V_R \rangle$, $\langle V_\Phi \rangle$ and $\langle V_Z \rangle$ all vary spatially in three dimensions, i.e. in ($R$, $\Phi$, $Z$), or ($X$, $Y$, $Z$). If the bulk motions are estimated by projecting the stars to a two-dimensional plane, the results are unlikely to be a good representation of the distribution in case where the bulk motions vary significantly in the projected dimension. For example, the mean $V_R$ of stars within bin ($R_1$ -- $R_2$, $Z_1$ -- $Z_2$) is a weighted mean of $\langle V_R \rangle$($\Phi$$\mid$$R_1$ -- $R_2$, $Z_1$ -- $Z_2$) with weights $n$($\Phi$$\mid$$R_1$ -- $R_2$, $Z_1$ -- $Z_2$), where $n$ is the stellar number density of the sample used. Samples of different $n$ could yield quite different results. This problem is naturally avoided by three-dimensional binning, as we have done here for the global sample.

\subsubsection{Selection Effects}
\label{sec_selection}

As we shall see in Section~\ref{sec_poplocal}, different populations seem to have different bulk motions. The differences in $\langle V_R \rangle$ and $\langle V_Z \rangle$ are relatively small, i.e. on the order of a few km~s$^{-1}$. However, $\langle V_\Phi \rangle$ has larger differences, $\sim$10~km~s$^{-1}$, between different populations. The global sample contains stars spanning a wide range of effective temperatures and metallicities, and thus suffer from selection effects. The selection effects in $\langle V_R \rangle$ and $\langle V_Z \rangle$ are less important, comparing with the systematic errors introduced by proper motions (from several to more than 10~km~s$^{-1}$; see Figs.~\ref{figvr1}-\ref{figvz2}). In contrast, the selection effects in $\langle V_\Phi \rangle$ are expected to be more significant. We do not correct for the selection effects in the current paper.

\subsubsection{The Solar Motions}
\label{sec_solar}

Bovy et al.~(2015) point out that bulk motions have a significant effect on the determination of the solar motion with respect to the LSR. With stars spanning from the solar vicinity out to $\sim$~5~kpc, they determine the azimuthal component of solar motion to be $V_\odot$ $\sim$ 24~km~s$^{-1}$, much larger than the previously locally-determined results. The discrepancy can be explained if the local Galactic disk has a $\sim$ 10~km~s$^{-1}$ bulk motion (Bovy et al.~2015). In this paper, we have calculated stellar velocities adopting a locally-determined set of solar motions, ($U_\odot$, $V_\odot$, $W_\odot$) = (7.01, 10.13, 4.95)~km~s$^{-1}$ (Huang et al.~2015). Here we do not aim to obtain a better set of solar motions, which may well vary depending on the spatial extension of stars used to determine them in view of the finding of Bovy et al. In any case, the bulk motions, $\langle V_R \rangle$, $\langle V_\Phi \rangle$ and $\langle V_Z \rangle$, reported here can be updated in the future by simply applying the corresponding offsets, $\delta U_\odot$, $\delta V_\odot$ and $\delta W_\odot$, between the future new and the current solar motions, as we approximately have $\delta (\langle V_R \rangle - U_\odot)$, $\delta (\langle V_\Phi \rangle - V_\odot)$ and $\delta (\langle V_Z \rangle - W_\odot)$ $\sim$ 0~km~s$^{-1}$ for our sample.

\subsection{The Bulk Motions}
\label{sec_bulk}

\begin{figure*}
\centering
\includegraphics[scale=0.50]{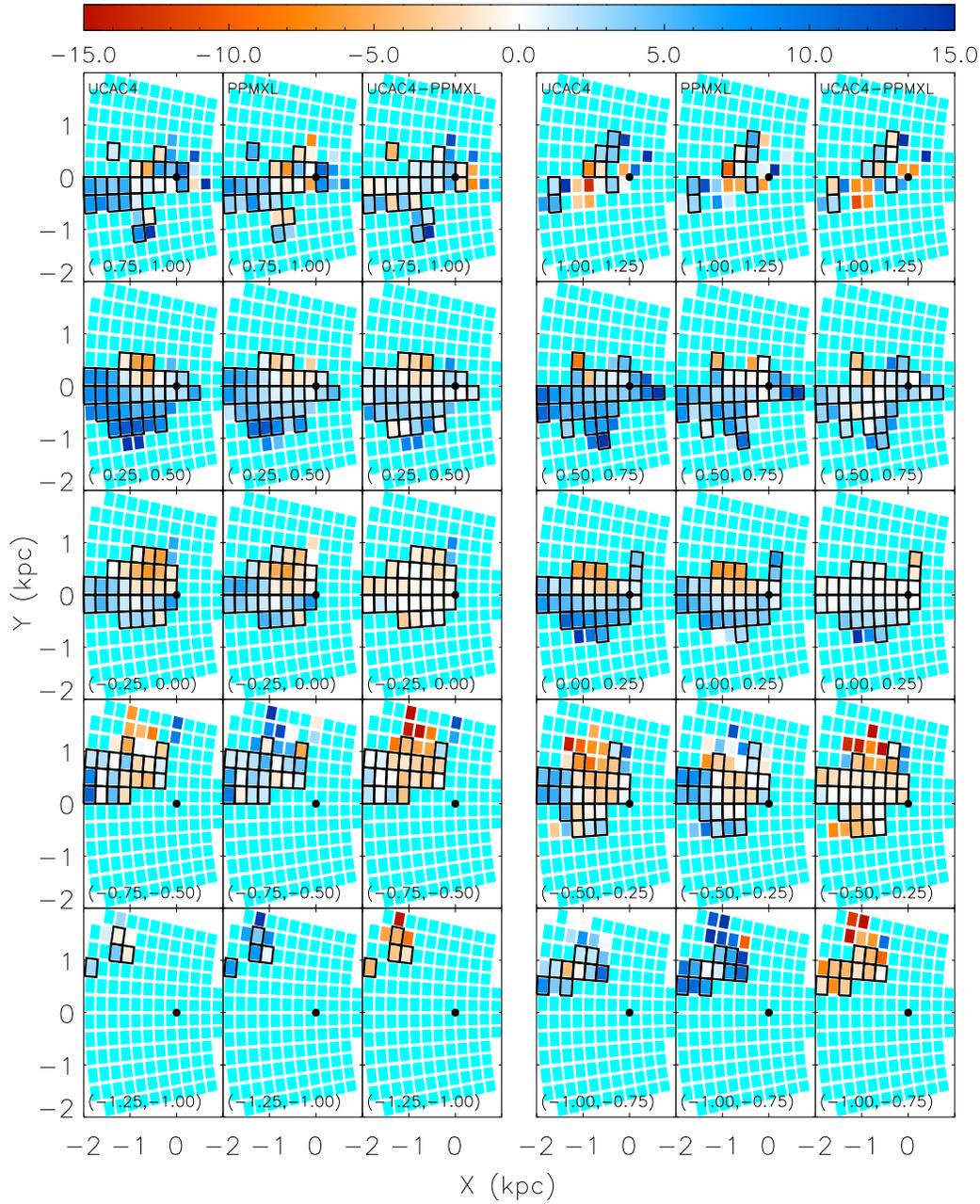}
\caption{Distributions of $\langle V_R \rangle $ obtained with either the UCAC4 or the PPMXL proper motions and their differences in units of~km~s$^{-1}$ viewed in the $X$--$Y$ plane. Boxes in black show bins where the differences are no larger than 5~km~s$^{-1}$. Numbers in the brackets show the ranges of $Z$ in units of kpc. The black dot denotes the location of the Sun.}
\label{figvr1}
\end{figure*}

\begin{figure*}
\centering
\includegraphics[scale=0.50]{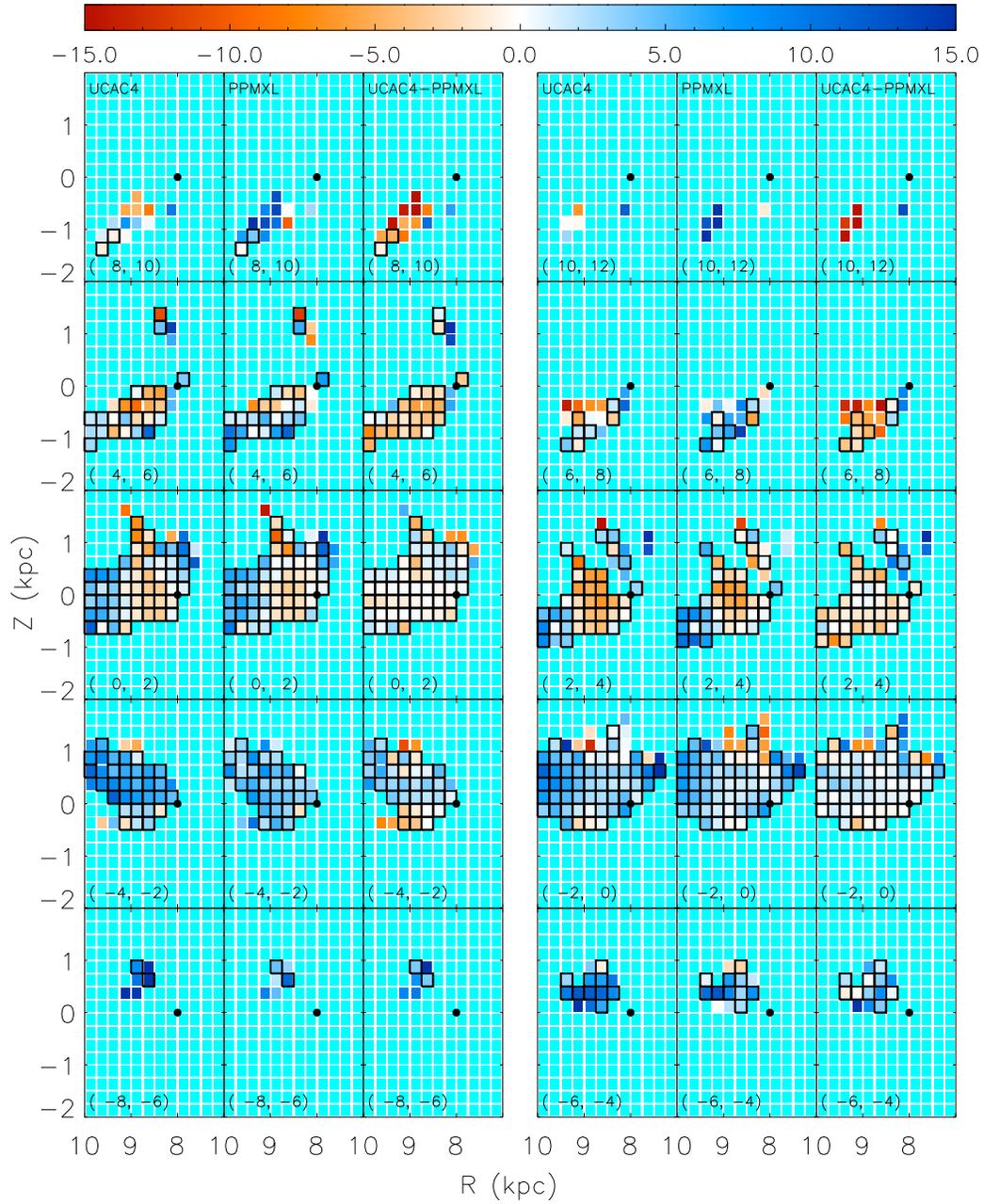}
\caption{Same as Fig.~\ref{figvr1} but for the $R$--$Z$ plane. Numbers in the brackets show the ranges of $\Phi$ in units of degrees.}
\label{figvr2}
\end{figure*}

\begin{figure*}
\centering
\includegraphics[scale=0.50]{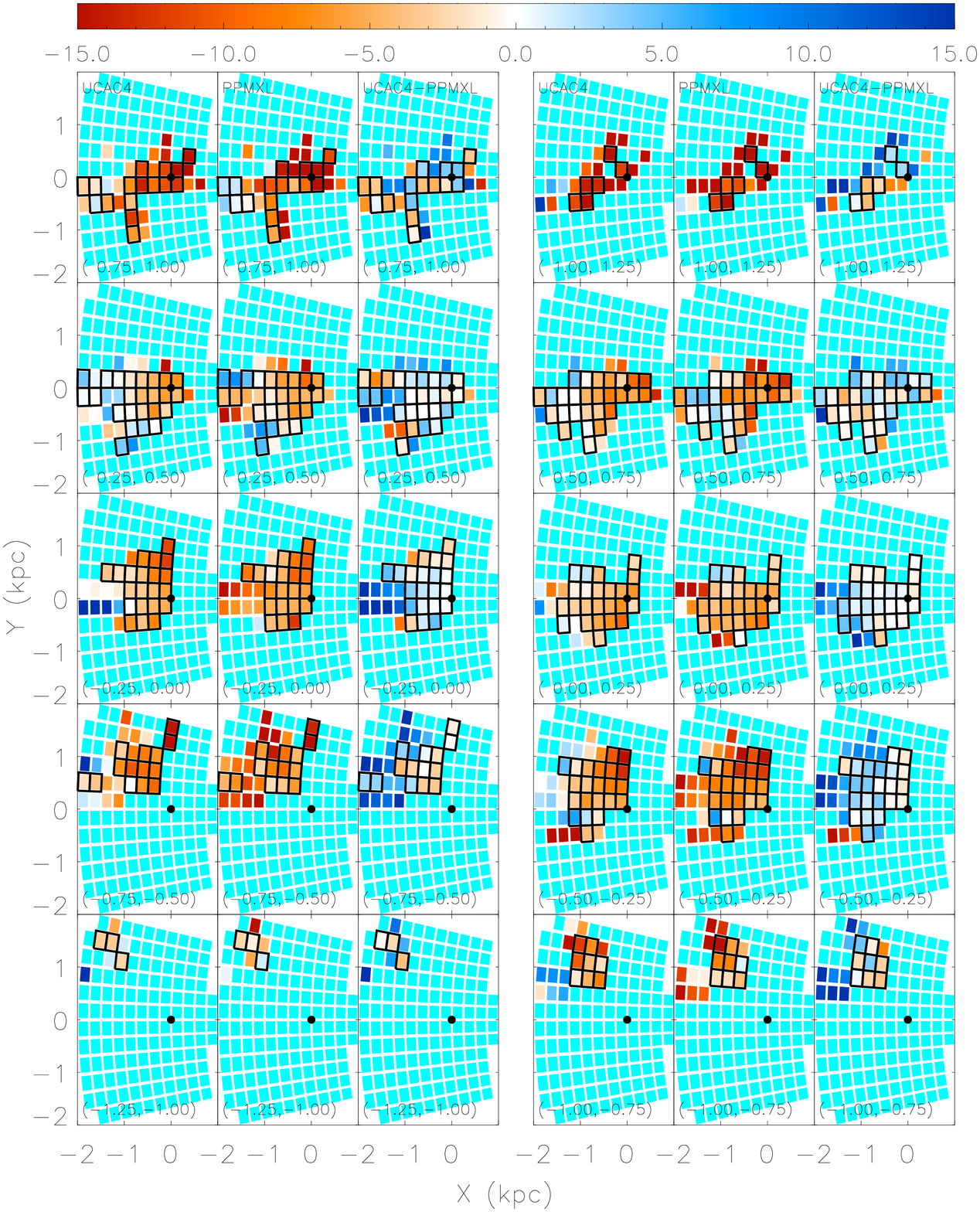}
\caption{Same as Fig.~\ref{figvr1} but for $\langle V_\Phi \rangle -$ 220~km~s$^{-1}$.}
\label{figvp1}
\end{figure*}

\begin{figure*}
\centering
\includegraphics[scale=0.50]{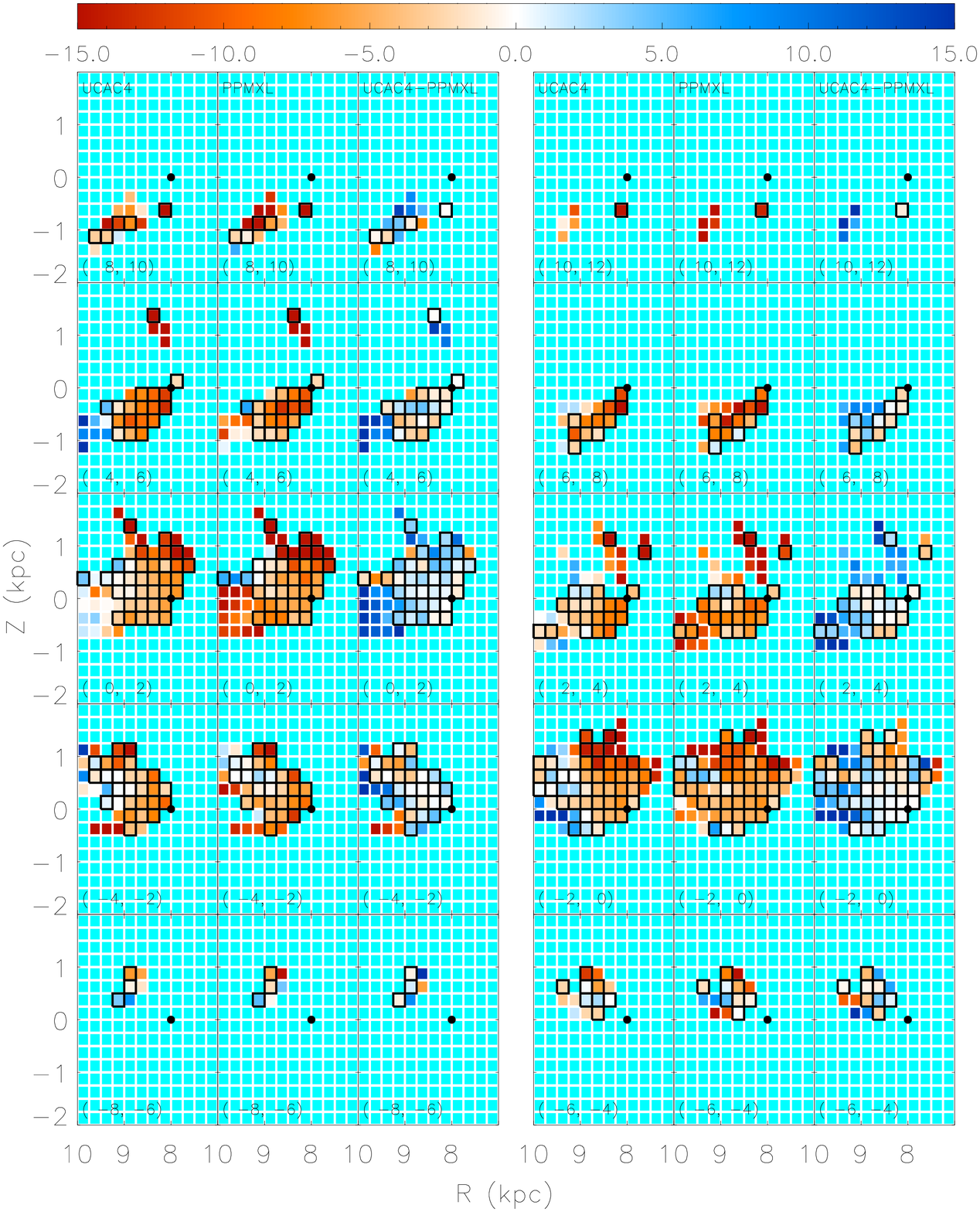}
\caption{Same as Fig.~\ref{figvr2} but for $\langle V_\Phi \rangle -$ 220~km~s$^{-1}$.}
\label{figvp2}
\end{figure*}

\begin{figure*}
\centering
\includegraphics[scale=0.50]{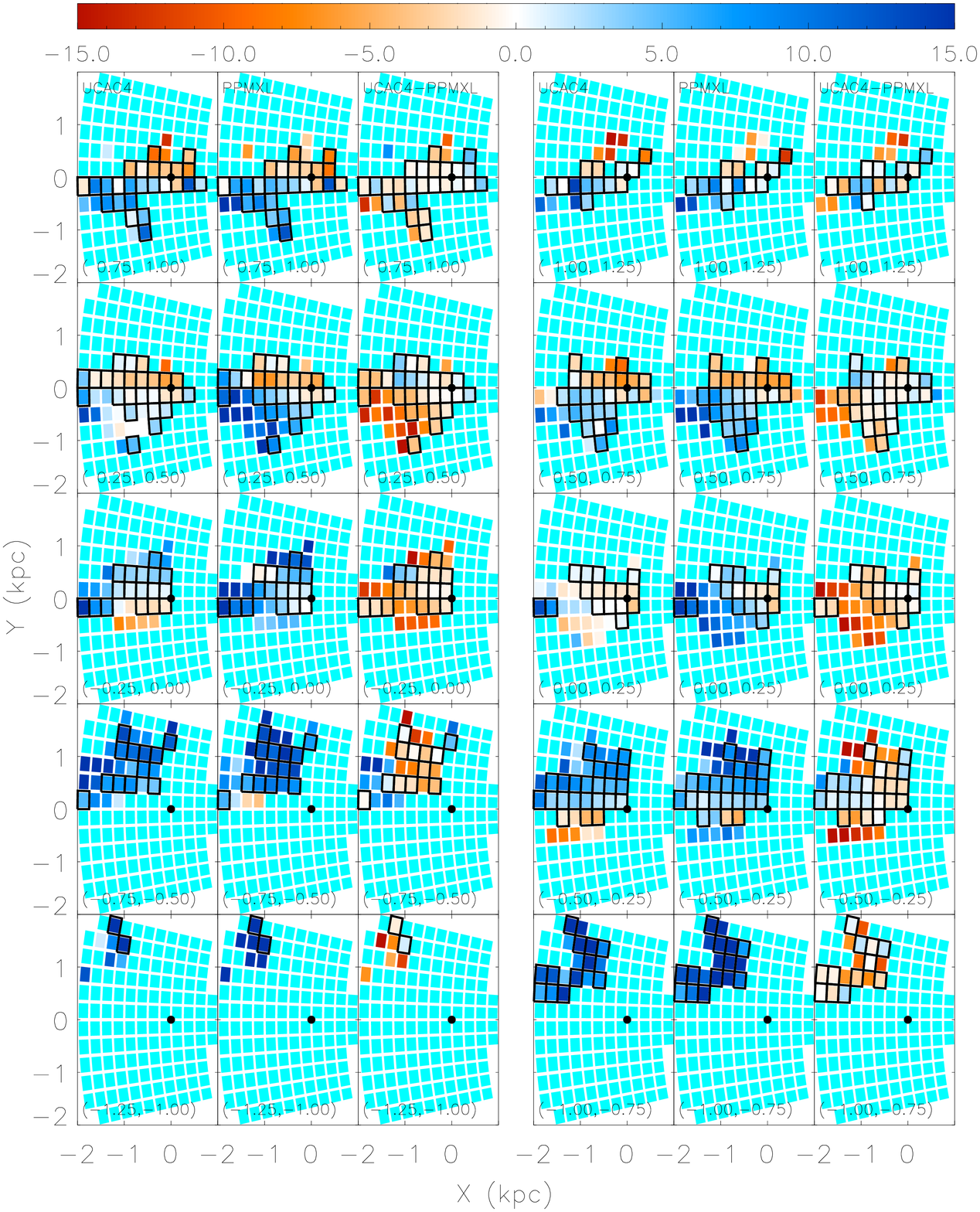}
\caption{Same as Fig.~\ref{figvr1} but for $\langle V_Z \rangle$.}
\label{figvz1}
\end{figure*}

\begin{figure*}
\centering
\includegraphics[scale=0.50]{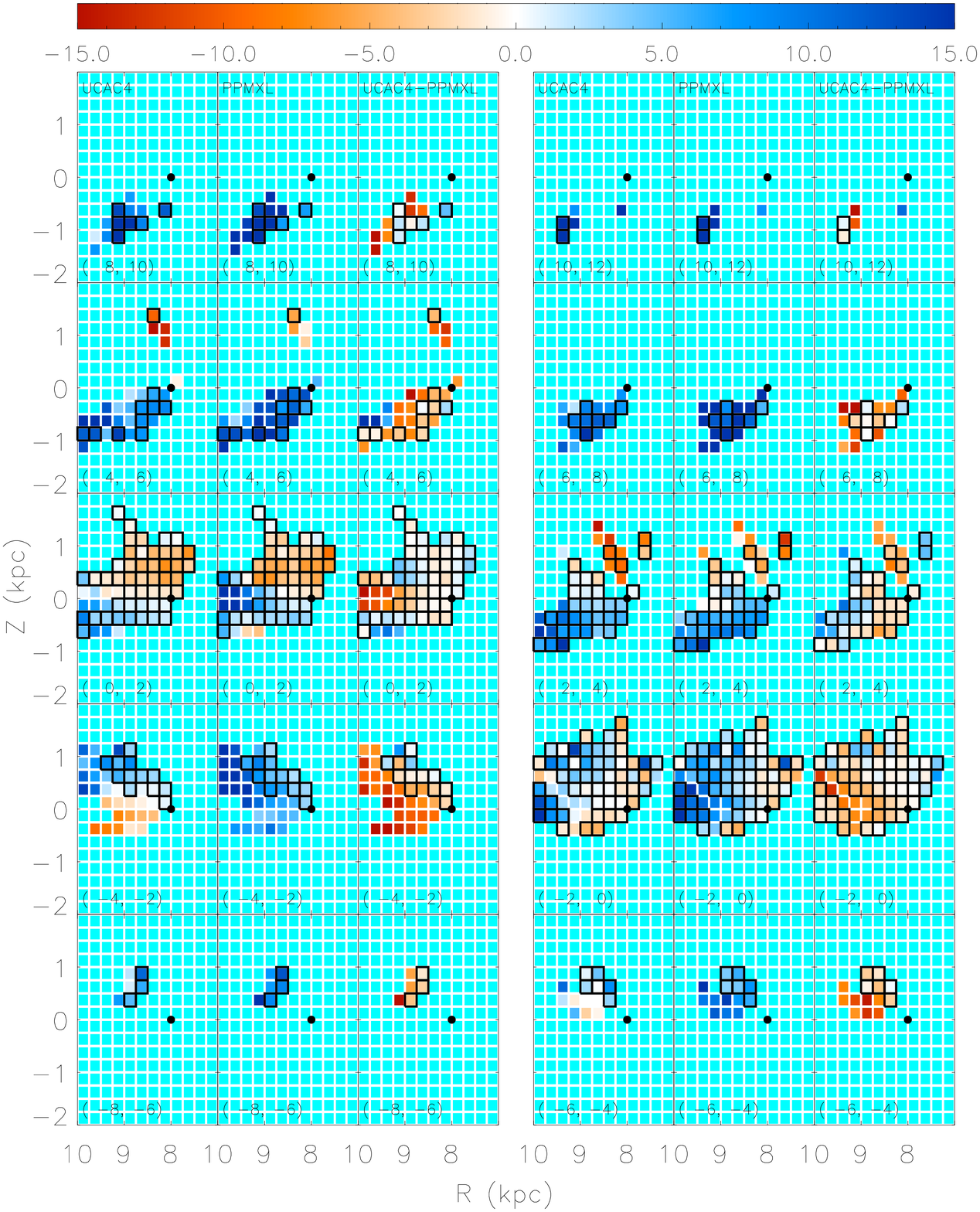}
\caption{Same as Fig.~\ref{figvr2} but for $\langle V_Z \rangle$.}
\label{figvz2}
\end{figure*}

\subsubsection{Radial Component $\langle V_R \rangle$}
\label{sec_vr}

Figs.~\ref{figvr1} and \ref{figvr2} show the maps of radial bulk motion $\langle V_R \rangle$. For most bins, the differences between the UCAC4 and PPMXL results are no larger than 5~km~s$^{-1}$. $\langle V_R \rangle$ is clearly nonzero, but ranges from $\sim$~$-$15~km~s$^{-1}$ to $\sim$~15~km~s$^{-1}$, consistent with the previous studies (e.g. Williams et al.~2013; Carlin et al.~2013). The majority parts of the disk mapped have outward bulk motions. However, inward motions dominate a large region near the Sun, i.e. the region of 8 $<$ $R$ $<$ 9~kpc, 0 $<$ $\Phi$ $<$ 6$^\circ$, $-$0.75 $<$ $Z$ $<$ 0.5~kpc, or roughly $-$1 $<$ $X$ $<$ 0~kpc, 0 $<$ $Y$ $<$ 1~kpc, $-$0.75 $<$ $Z$ $<$ 0.5~kpc.

\subsubsection{Azimuthal Component $\langle V_\Phi \rangle$}
\label{sec_vp}

Figs.~\ref{figvp1} and \ref{figvp2} show the maps of azimuthal bulk motion $\langle V_\Phi \rangle$, from which we have subtracted a constant circular speed of $V_c$ = 220~km~s$^{-1}$ for clarity. The UCAC4 and PPMXL proper motions give consistent results for most bins, with differences no larger than 5 km s$^{-1}$. The Galactic disk is dominated by negative values of $\langle V_\Phi \rangle - V_c$, which decreases systematically as one moves away from the mid-plane. Near $Z$ $\sim$ 0~kpc, the stars lag behind the circular speed by $\sim$~8~km~s$^{-1}$; but at $\mid$$Z$$\mid$ $\sim$ 1~kpc, the stars have $\langle V_\Phi \rangle - V_c$ $\lesssim$ $-$15~km~s$^{-1}$. Similar trends are also unveiled by Williams et al.~(2013). As mentioned in Section~\ref{sec_selection}, $\langle V_\Phi \rangle$ could be affected by some selection effects. The later-type stars, which are expected to have larger asymmetric drifts, are concentrated close to the mid-plane. Thus, $\langle V_\Phi \rangle$ at larger values of $\mid$$Z$$\mid$ could be slightly over-estimated due to the selection effects.
%It is worth noting that the non-zero values of $\langle V_\Phi \rangle$ $-$ $V_c$ arise from the combined effects of non-axisymmetries and asymmetric drift, which we do not attempt to distinguish in this paper.

\subsubsection{Vertical Component $\langle V_Z \rangle$}
\label{sec_vz}

Figs.~\ref{figvz1} and \ref{figvz2} show the maps of vertical bulk motion $\langle V_Z \rangle$. Similar to $\langle V_R \rangle$, $\langle V_Z \rangle$ ranges approximately from $-$15 to 15~km~s$^{-1}$, again in agreement with the previous studies (Widrow et al.~2012; Williams et al.~2013; Carlin et al.~2013). However, the vertical bulk motions are more severely affected by the systematic errors in proper motions, and there are more bins where UCAC4 and PPMXL give discrepant values. Williams et al.~(2013) and Carlin et al.~(2013) have found that the Galactic disk exterior to the solar position is compressing, i.e. stars above the mid-plane move downwards while those below upwards. Our results confirm that the disk sector of 0 $< \Phi <$ 6$^\circ$ is indeed compressing, but that of $-$4 $<$ $\Phi$ $<$ 0$^\circ$ does not. The UCAC4 results even show that the disk sector of $-$4 $<$ $\Phi$ $<$ $-$2$^\circ$ is probably rarefying. Stars within 9 $< R <$ 10~kpc and $-$2 $<$ $\Phi$ $<$ 0$^\circ$, above or below the disk mid-plane regardless, are found to be co-moving upwards, indicative of a bending-mode perturbation. More discussions on the vertical perturbations can be found in Section~\ref{sec_vertical}.

\subsection{Giant Stellar Flows in the Disk}
\label{sec_flow}

\begin{figure*}
\centering
\includegraphics[scale=0.36]{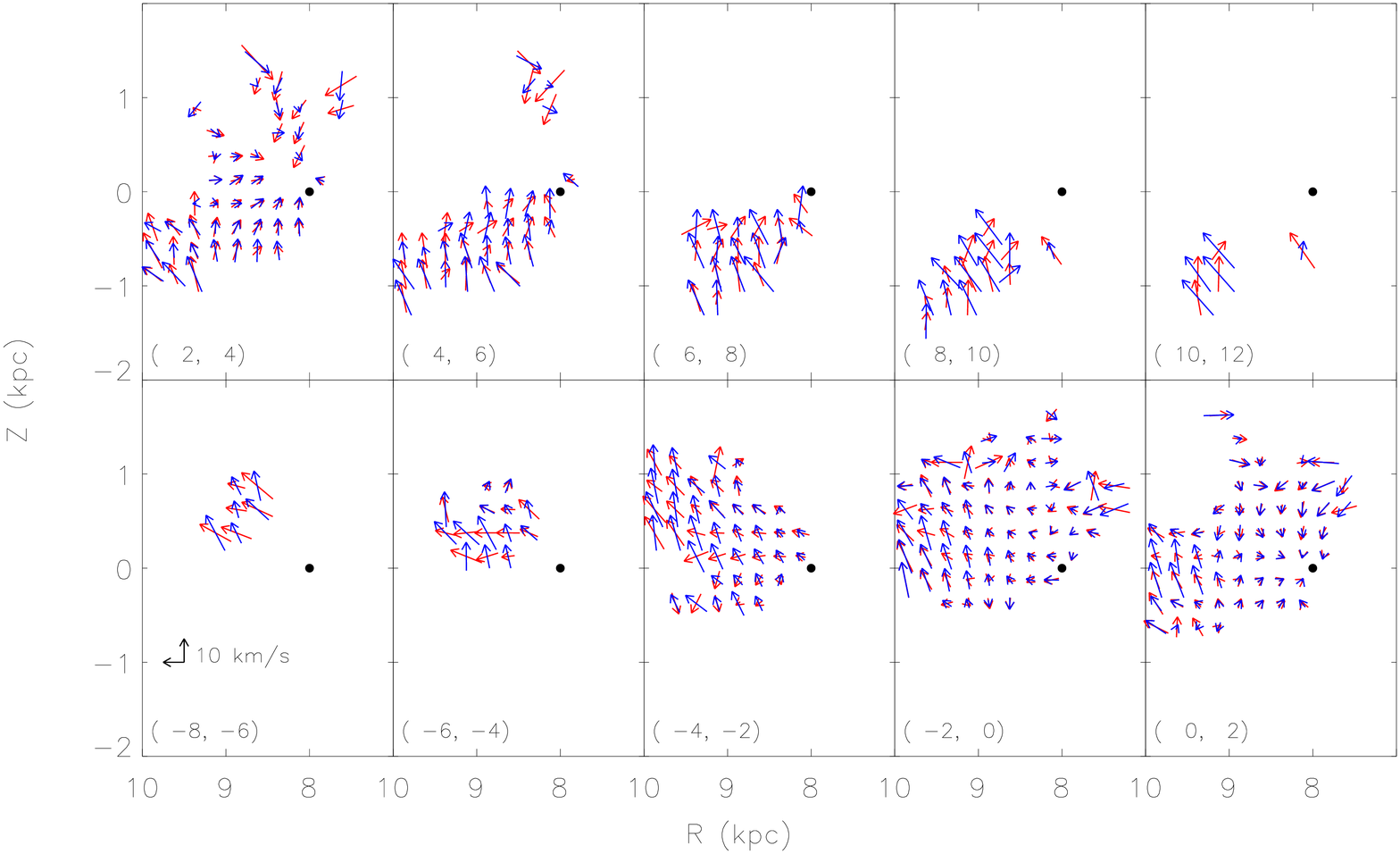}
\caption{Distributions of the magnitude and direction of the total velocity $\sqrt{\langle V_R \rangle^2 + \langle V_Z \rangle^2}$ in the $R$--$Z$ plane. The arrows indicate the flow directions and their lengths are proportional to the total velocities. Red arrows are calculated with the UCAC4 proper motions, whereas those in blue are results from the PPMXL. Numbers in the brackets show the ranges of $\Phi$ in units of degrees. The black dot denotes the location of the Sun.}
\label{figvv1}
\end{figure*}

\begin{figure*}
\centering
\includegraphics[scale=0.36]{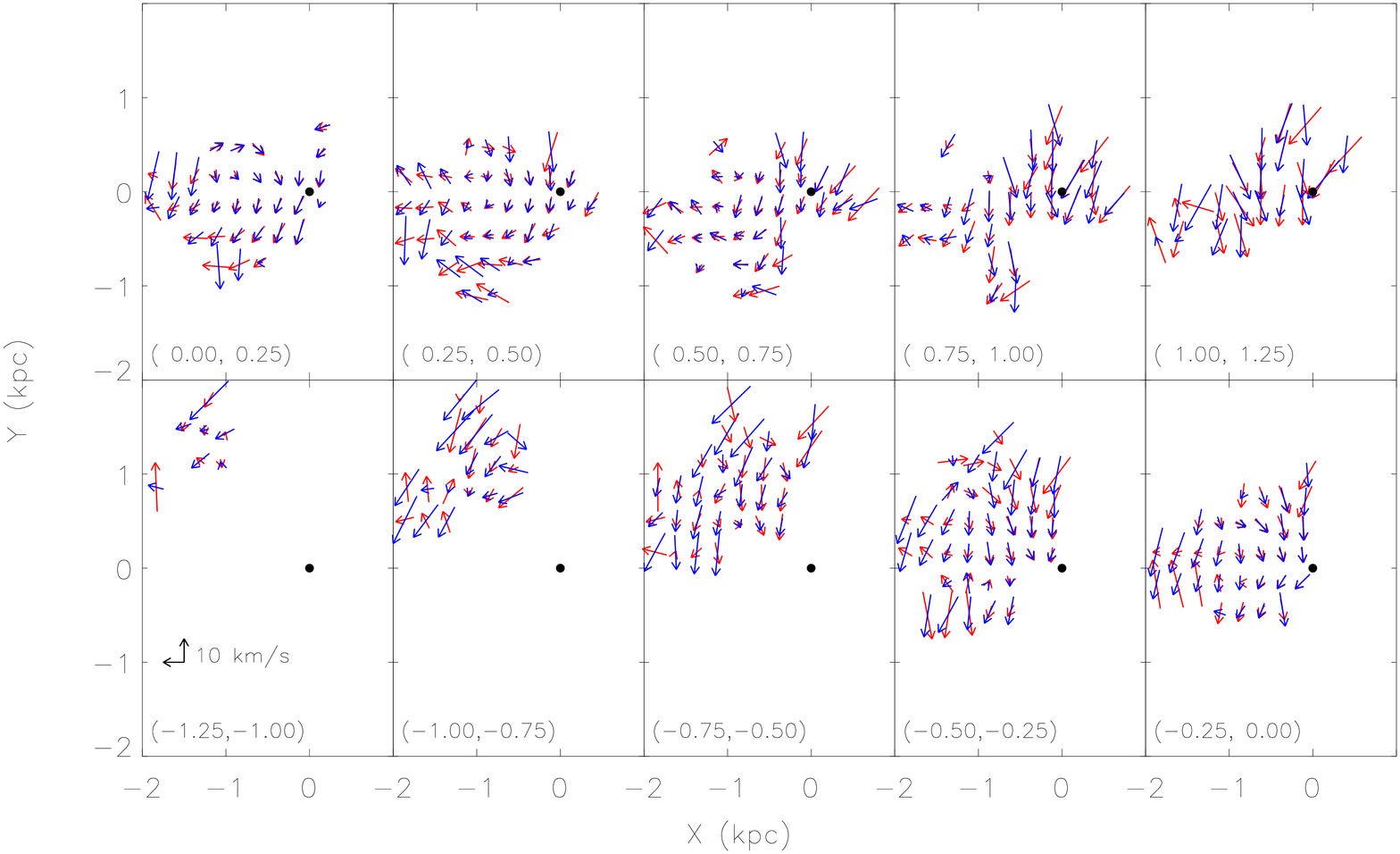}
\caption{Same as Fig.~\ref{figvv1} but for the total velocity $\sqrt{\langle V_R \rangle^2 + (\langle V_\Phi \rangle - V_c)^2}$ in the $X$--$Y$ plane. Numbers in the brackets show the ranges of $Z$ in units of kpc.}
\label{figvv2}
\end{figure*}

The bulk motion maps unveil spatially coherent, kpc-scale flows in the disk, as can be seen in Figs.~\ref{figvv1} and \ref{figvv2}. At most locations, the UCAC4 and PPMXL proper motions give compatible results in terms of both speeds and directions. Fig.~\ref{figvv1} displays $\langle V_R \rangle$ and $\langle V_Z \rangle$ as vectors in the $R$--$Z$ plane. From $\Phi$ = $-$8$^\circ$ to $\Phi$ = $-$2$^\circ$, a giant flow, of kpc scales and typical velocities $\sqrt{\langle V_R \rangle^2 + \langle V_Z \rangle^2}$ $\gtrsim$ 10~km~s$^{-1}$, streams from the solar neighborhood out- and up-wards. The nearer parts of the flow ($R \leq$ 9~kpc) become weaker in terms of typical velocity in the sector $-$2 $<$ $\Phi$ $<$  0$^\circ$ and the direction is reversed between $\Phi$ = 0 and 2$^\circ$: the flow now runs from outside into the solar neighborhood. For $\Phi$ $>$ 2$^\circ$, we see the flow increases its speed back to $\gtrsim$~10~km~s$^{-1}$, reaching a maximum of $\sim$~20~km~s$^{-1}$.

Fig.~\ref{figvv2}, which plots $\langle V_R \rangle$ and $\langle V_\Phi \rangle - V_c$ in the $X$--$Y$ plane, displays the giant flow from another perspective. Due to the large asymmetric drifts, the giant flow lags behind the circular speed by $\ge$~10~km~s$^{-1}$ at $Z$ $<$ $-$0.5~kpc or $Z$ $>$ 0.75~kpc. Near the mid-plane, one sees an incomplete, clockwise vortex structure around $-$1 $<$ $X$ $<$ 0~kpc and $-$1 $<$ $Y$ $<$ 0.5~kpc. It has a diameter of $\sim$~1~kpc and a typical velocity $\sqrt{\langle V_R \rangle^2 + (\langle V_\Phi \rangle - V_c)^2}$ $\lesssim$~10~km~s$^{-1}$. This structure is most obvious between 0 $<$ $Z$ $<$ 0.5~kpc.

\subsection{Vertical Perturbations}
\label{sec_vertical}

The vertical kinematics has drawn considerable interest. Here we use $\langle V_Z \rangle$($Z$$\mid$$R$, $\Phi$) to denote the vertical bulk motion $\langle V_Z \rangle$ as a function of height $Z$ at a specific location ($R$, $\Phi$) in the disk plane. It has been shown that $\langle V_Z\rangle$($Z$$\mid$$R$, $\Phi$) can be decomposed into the bending and breathing modes plus the higher-order modes (Mathur~1990; Weinberg~1991; Widrow et al.~2014). The bending mode describes a situation where stars below and above the disk mid-plane move synchronously upwards or downwards, whereas the breathing mode refers to a scenario where stars below and above the mid-plane move in opposite vertical directions. The disk is said to be in compression when stars above the mid-plane move downwards and those below upwards, or in rarefaction vice versa.

\begin{figure*}
\centering
\includegraphics[scale=0.62]{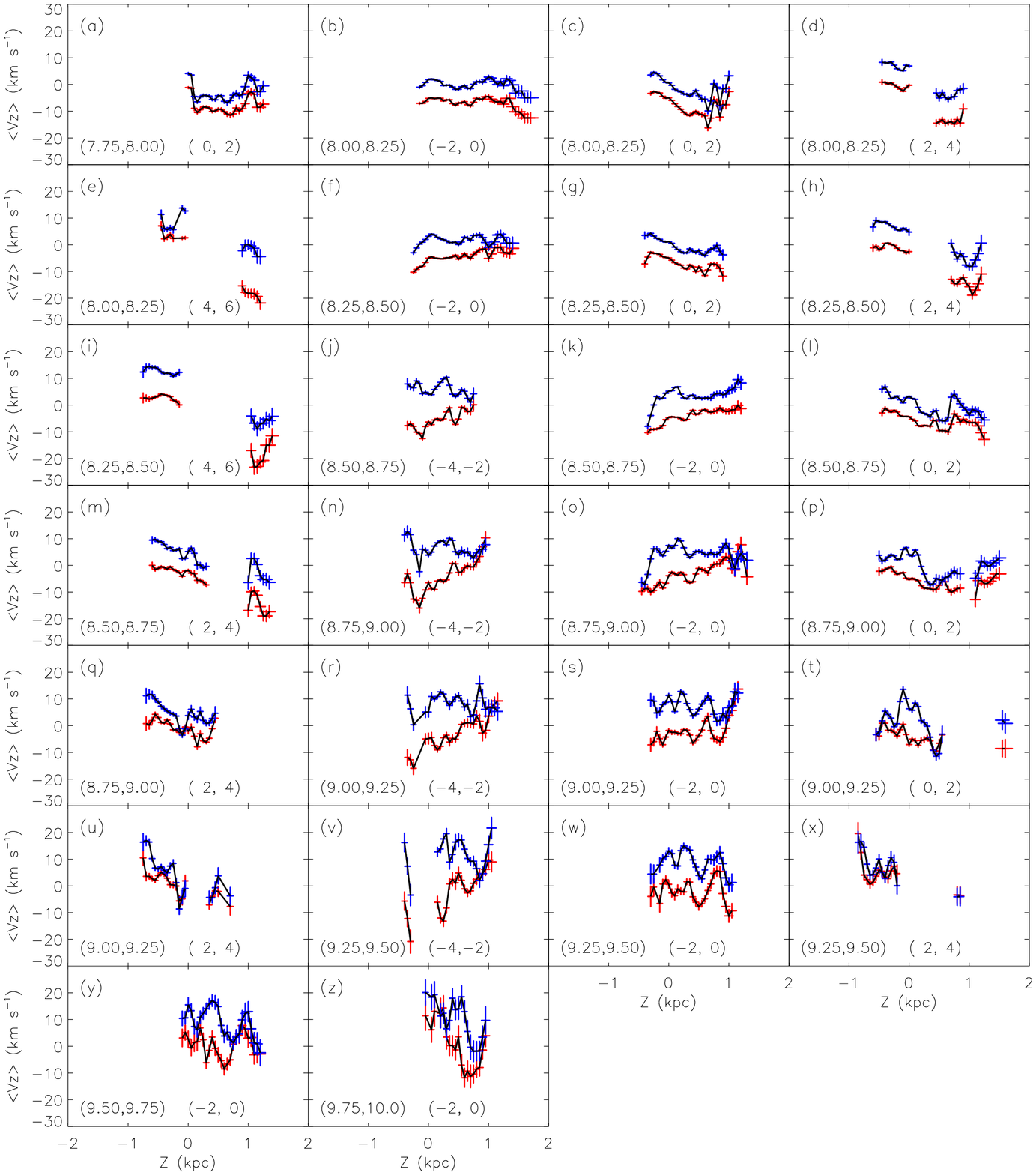}
\caption{Profiles of $\langle V_Z \rangle$ as a function of $Z$. In each panel, the first brackets give the range of $R$ in units of kpc, and the second that of $\Phi$ in units of degrees. Curves with red error bars have used the UCAC4 proper motions, while those with blue ones have used the PPMXL proper motions. The vertical error bars denote the 1-$\sigma$ random errors of $\langle V_Z \rangle$, whereas the horizontal ones are set to the binsize $\Delta Z$. The UCAC4/PPMXL results have been shifted by $+$/$-$5~km~s$^{-1}$, respectively, for clarity.}
\label{figvz1d}
\end{figure*}

To better display $\langle V_Z \rangle$($Z$$\mid$$R$, $\Phi$), we keep the binsize in the disk plane $\Delta R$ = 250~pc and $\Delta \Phi$ = 2$^\circ$ unchanged but increase the sampling frequency in height $Z$. Bins in $Z$ are now spaced every 50~pc and have a height $Z$ $\times$ 15\% (corresponding to the random errors) but no smaller than 100~pc. As before, $\langle V_Z \rangle$ is estimated by the median $V_Z$ of stars in each bin. The curves of $\langle V_Z \rangle$($Z$$\mid$$R$, $\Phi$) are displayed in Fig.~\ref{figvz1d} for cases where there are at least 10 data points and the range of heights spans at least 1~kpc.

\subsubsection{Bending and Breathing Modes}
\label{sec_bb}

Despite the complicated behavior of $\langle V_Z \rangle$($Z$$\mid$$R$, $\Phi$), we find that a simple linear fit can be applied to roughly describe their trends (see also Widrow et al.~2014), i.e.
\begin{equation}
\langle V_Z \rangle(Z\mid R, \Phi) = A(R, \Phi) \times Z + B(R, \Phi)
\end{equation}
where the intersection $B$, in units of~km~s$^{-1}$, parameterizes the bending mode (positive for upward and negative for downward co-movement) and the slope $A$, in units of~km~s$^{-1}$~kpc$^{-1}$, parameterizes the breathing mode (positive for rarefaction and negative for compression).

\begin{figure*}
\centering
\includegraphics[scale=0.8]{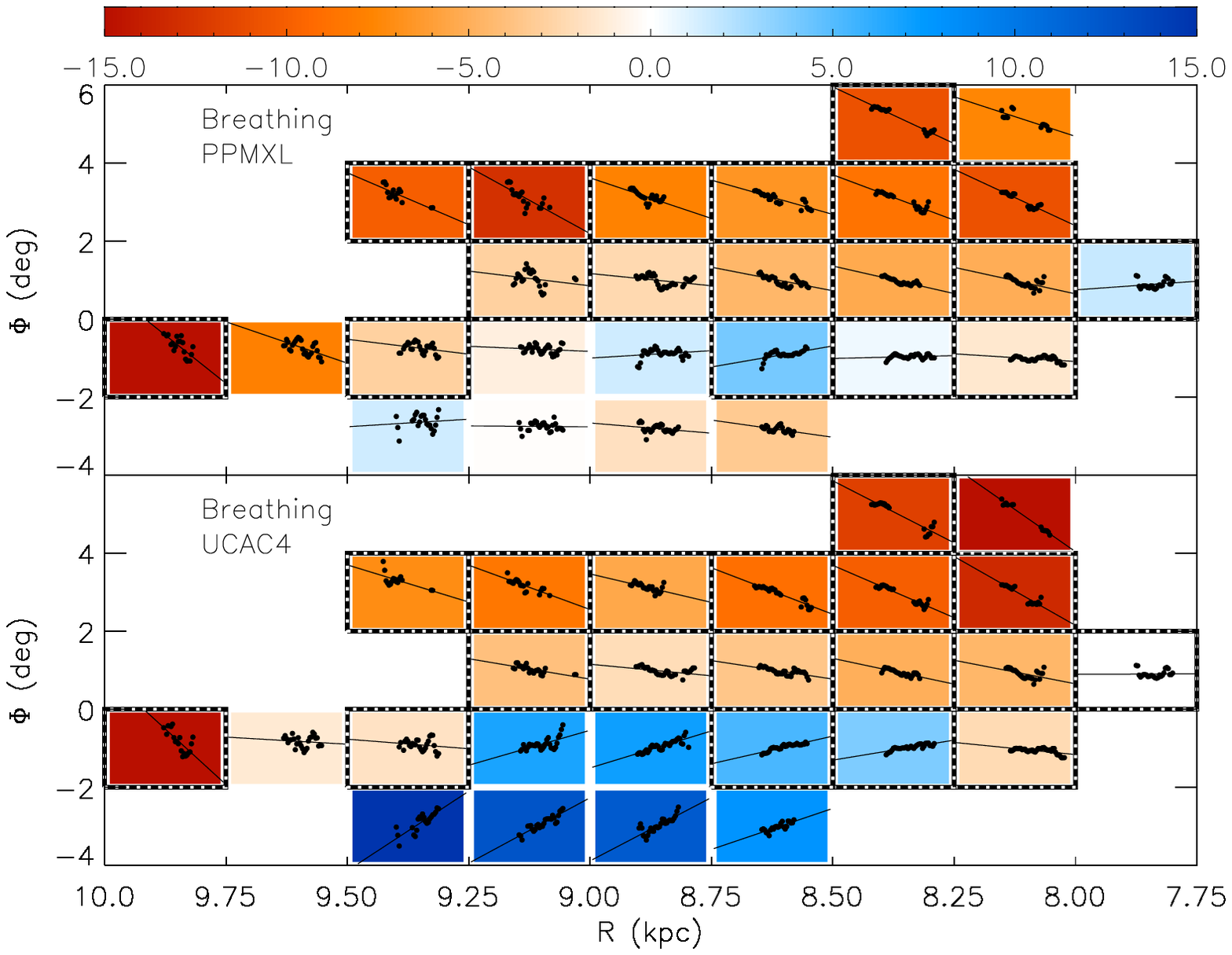}
\caption{Breathing mode parameter $A$ in units of~km~s$^{-1}$~kpc$^{-1}$ (colorscale) derived with the UCAC4 (lower panel) or the PPMXL (upper panel) proper motions. Each sub-panel plots $\langle V_Z \rangle$ as a function of $Z$ (points) along with a linear fit to the data (straight line). The horizontal axis of each sub-panel ranges from $-$2 to 2~kpc and the vertical axis from $-$30~km~s$^{-1}$ to 30~km~s$^{-1}$. Sub-panels enclosed in dotted black boxes show bins where the differences of $A$ yielded by the UCAC4 and PPMXL proper motions differ by less than 5~km~s$^{-1}$~kpc$^{-1}$.}
\label{figbb1}
\end{figure*}

The distributions of the breathing-mode parameter $A$ is displayed in Fig.~\ref{figbb1}. The figure once again confirms the compression and rarefaction patterns already described in Section~\ref{sec_vz} but denoted here by reddish and blueish cells. Typical values of $A$ range from $\sim$~$-$15~km~s$^{-1}$~kpc$^{-1}$ to $\sim$~15~km~s$^{-1}$~kpc$^{-1}$ (with the UCAC4 proper motions) or to $\sim$~0~km~s$^{-1}$~kpc$^{-1}$ (with the PPMXL proper motions). Out of 26 cells in total, 18 have the UCAC4 or PPMXL results consistent with each other within 5~km~s$^{-1}$~kpc$^{-1}$. Results from both sets of proper motions reveal a smooth, gradual change of $A$ on~kpc scales across the disk plane. The gradient is estimated to be roughly 18~km~s$^{-1}$~kpc$^{-2}$.

\begin{figure*}
\centering
\includegraphics[scale=0.8]{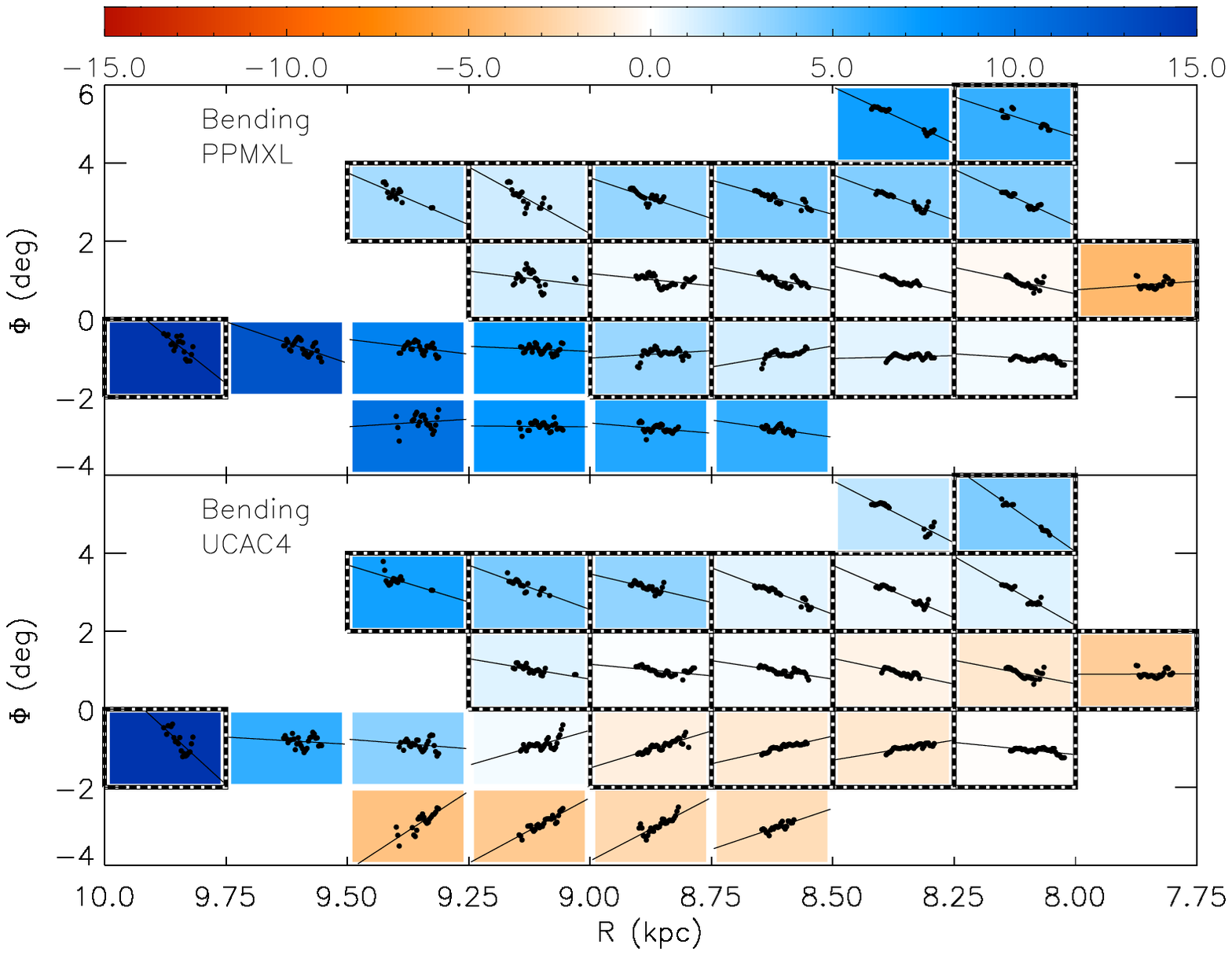}
\caption{Same as Fig.~\ref{figbb1} but for bending mode parameter $B$ in units of~km~s$^{-1}$. Sub-panels enclosed in dotted black boxes show bins where the UCAC4 and PPMXL values of $B$ from UCAC4 and PPMXL differ by less than 5~km~s$^{-1}$.}
\label{figbb2}
\end{figure*}

Fig.~\ref{figbb2} shows the distribution of bend-mode parameter $B$, also found to possess smooth spatial variations. In the case of UCAC4 proper motions, the Galactic disk is found to have both upward or downward bending modes. However, with the PPMXL proper motions, we find that almost the whole region investigated has an upward bending mode only, except for one cell in the solar vicinity with $B$~$\sim$~$-$5~km~s$^{-1}$. Still, more than two thirds of the cells have the UCAC4 or PPMXL results consistent with each other within 5~km~s$^{-1}$. A gradient of $\sim$~15~km~s$^{-1}$~kpc$^{-1}$ is estimated for the bending mode parameter.

\subsubsection{Breaks and Ripples}
\label{sec_br}

In addition to the bending and breathing modes, the vertical bulk motion exhibits some higher-order perturbations, such as breaks or ripples along the $\langle V_Z \rangle$($Z$$\mid$$R$, $\Phi$) curves. A representative example of the breaks can be seen in Panel (l) of Fig.~\ref{figvz1d} (8.5c $<$ $R$ $<$ 8.75~kpc, 0 $<$ $\Phi$ $<$ 2$^\circ$). $\langle V_Z \rangle$ first decreases smoothly from $Z$ = $-$0.5~kpc to $Z$ = 0.6~kpc, after which it jumps sharply by $\sim$ 10~km~s$^{-1}$ within 0.2~kpc and then comes back again to a smooth gradual decrease with $Z$ at $Z$ $>$ 0.8~kpc. The entire curve is broken by the two turning points at $Z$ = 0.6 and 0.8~kpc into three segments, each well approximated by a linear function. Similar breaks can also be found in Panels (a, c, f).

\begin{figure*}
\centering
\includegraphics[scale=0.7]{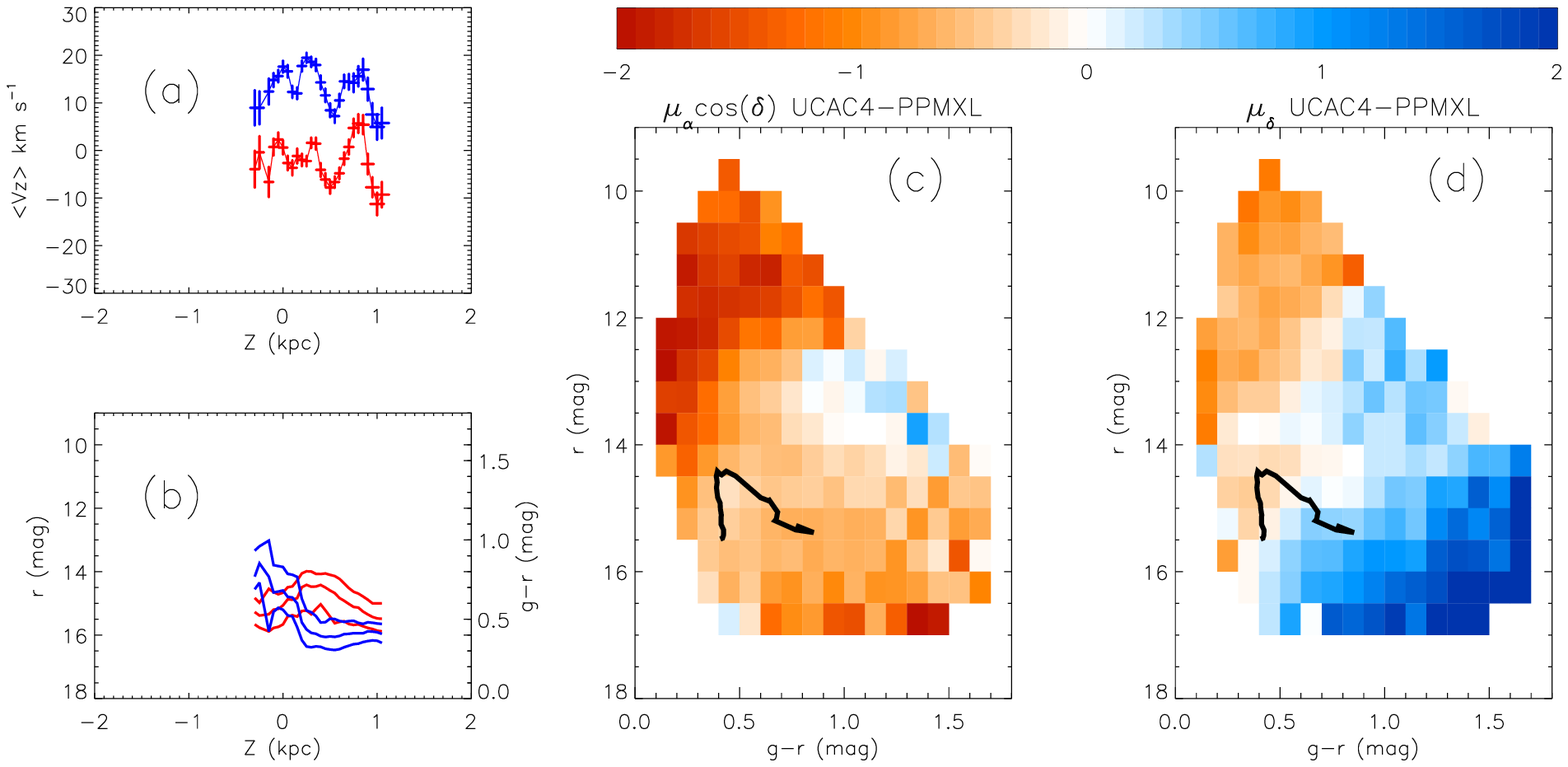}
\caption{Panel (a): $\langle V_Z \rangle$ as a function of $Z$ for 9.25 $<$ $R$ $<$ 9.5~kpc and $-$2 $<$ $\Phi$ $<$ 0$^\circ$, same as for Panel (w) of Fig.~\ref{figvz1d}. Panel (b): Distribution of $g-r$ colors (blue curves) and $r$ magnitudes (red curves) as a function of $Z$, where the central lines denote the median values of $g-r$ or $r$ and 25\% stars have values above/below the upper/lower lines. Panels (c, d): Locus of the median values of $g-r$ and $r$ in the ($g-r$, $r$) Hess diagram. The colorscale denotes the median differences between the UCAC4 and PPMXL proper motions, same as the colorscale in Fig.~\ref{figsys}.}
\label{figck}
\end{figure*}

The second kind of higher-order perturbations are characterized by the high-frequency ripples in the $\langle V_Z \rangle$($Z$$\mid$$R$, $\Phi$) curves, especially at large distances from the Sun, e.g. Panels (o-z) of Fig.~\ref{figvz1d}. The ripples have peak-to-peak amplitudes $A_{\rm pp}$ up to $\sim$ 10~km~s$^{-1}$ and frequencies $k_Z$ of roughly 2--5~kpc$^{-1}$. Similar ripples have been found by Widrow et al.~(2012), with comparable peak-to-peak amplitudes and frequencies. One may question whether these ripples arise from the unaccounted for color- and magnitude-dependent systematic errors in the proper motions. However, we believe that those ripples are probably real, as the color and magnitude distributions of stars vary smoothly with $Z$. An example is given in Fig.~\ref{figck}, which shows the distributions of $g-r$ colors and $r$-band magnitudes of stars of 9.25 $<$ $R$ $<$ 9.5~kpc and $-$2 $<$ $\Phi$ $<$ 2$^\circ$ [corresponding to Panel (w) of Fig.~\ref{figvz1d}]. It can be seen that both $g-r$ and $r$ vary smoothly with $Z$, with no clear features corresponding to the peaks or dips of $\langle V_Z \rangle$ at $Z$ = 0, 0.1, 0.3, 0.5 or 0.8~kpc. A small feature in $g-r$ at $Z$ = $-$0.2~kpc is possibly related to a small dip at the same location in $\langle V_Z \rangle$ obtained with the UCAC4 proper motions. Besides, the locus of $g-r$ and $r$ median values in the ($g-r$, $r$) Hess diagram occupies a region where the differences of $\mu_\alpha$cos$\delta$ between UCAC4 and PPMXL are almost constant while the differences of $\mu_\delta$ vary smoothly from 3 to $-$2~mas~yr$^{-1}$. Although we do not know the systematic errors of UCAC4 or PPMXL proper motions themselves (we only know their differences), it is unlikely that they all vary quickly along the locus while at the same time generating smoothly varying differences. Thus it seems to us that the systematic errors of proper motions may not be responsible for the observed high-frequency ripples along the $\langle V_Z \rangle$($Z$$\mid$$R$, $\Phi$) curves.

\section{Bulk Motions of the Local Stellar Populations}
\label{sec_poplocal}

It is an interesting problem whether different stellar populations share the same bulk motions. Smith et al.~(2012) and Widrow et al.~(2012) show that there are differences of up to $\sim$~5~km~s$^{-1}$ between stars of different metallicities or colors. Such a study is not feasible for the global sample, as the potential color-dependent systematic errors in proper motions may generate false bulk motion differences amongst populations of different colors, especially at large distances. We thus restrict ourselves to the local sample, which contains 5,443 FGK stars of distances $\leq$~150~pc from the Sun (see Section~\ref{sec_lcsam}). At this distance limit, a  1~mas~yr$^{-1}$ error in the proper motion corresponds to a $\sim$ 0.7~km~s$^{-1}$ error in the tangential velocity perpendicular to the line of sight.

\begin{figure}
\centering
\includegraphics[scale=0.8]{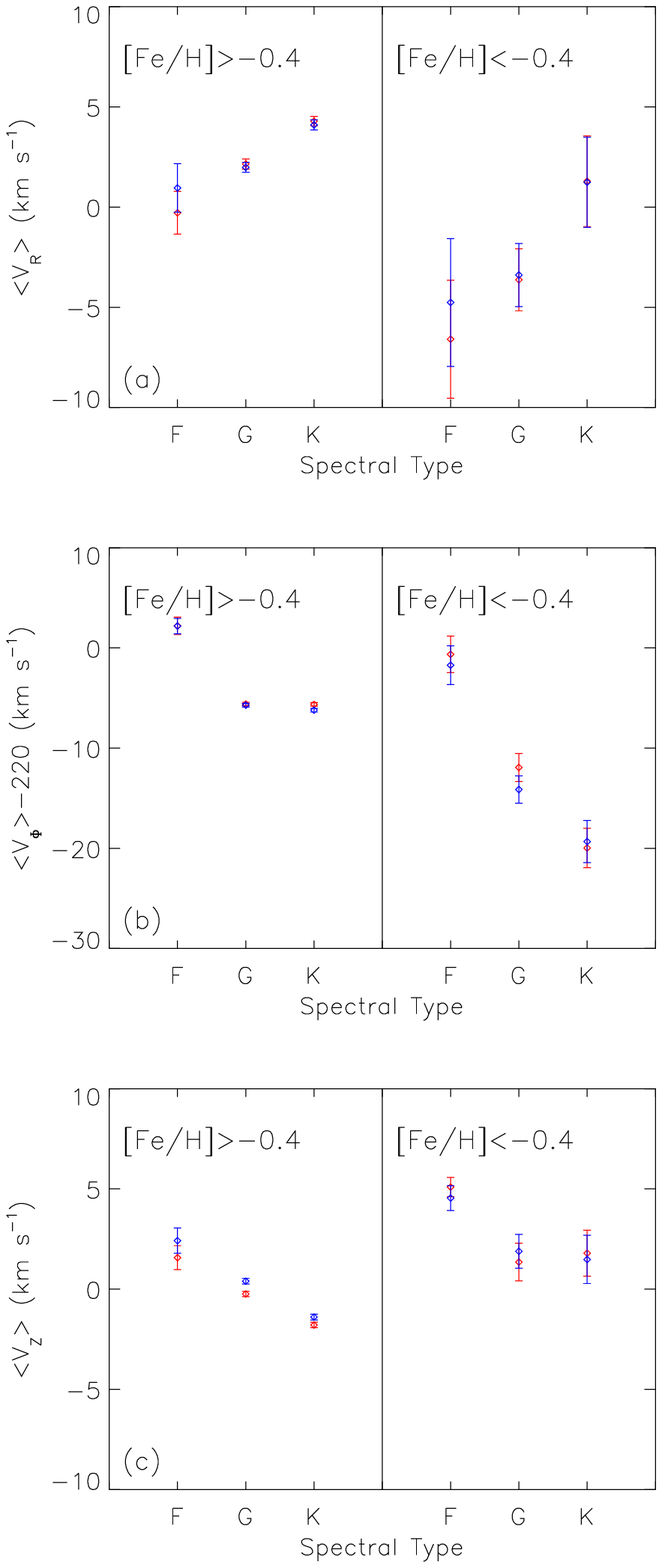}
\caption{$\langle V_R \rangle$, $\langle V_\Phi \rangle$ and $\langle V_Z \rangle$ for different populations of the local sample. The red symbols are calculated with the UCAC4 proper motions while those in blue with the PPMXL values.}
\label{figpop}
\end{figure}

The local sample is divided into a metal-rich group ([Fe/H] $>$ $-$0.4~dex) and a metal-poor group ([Fe/H] $<$ $-$0.4~dex). Each of the two groups is further divided into three based on the spectral-types, the F-type (6,000 $<$ $T_{\rm eff}$ $<$ 6,800~K), G-type (5,000 $<$ $T_{\rm eff}$ $<$ 6,000~K ) and K-type (4,200 $<$ $T_{\rm eff}$ $<$ 5,000~K). $\langle V_R \rangle$, $\langle V_\Phi \rangle$ and $\langle V_Z \rangle$ are estimated by the median velocities of stars of each population, and the results can be found in Table~1 and Fig.~\ref{figpop}.

The differences between UCAC4 and PPMXL results for each population are smaller than $\sim$~1~km~s$^{-1}$, or within the random errors. However, the differences between different populations, either in $\langle V_R \rangle$, $\langle V_\Phi \rangle$ or $\langle V_Z \rangle$, are well beyond the errors, at levels of high significance. The variations of azimuthal bulk motions are the most obvious, with $\langle V_\Phi \rangle -$ 220~km~s$^{-1}$ ranging from $\sim$~2~km~s$^{-1}$ for the metal-rich F-type stars to $\sim$~$-$20~km~s$^{-1}$ for the metal-poor K-type stars. This is what expected, considering the larger velocity dispersions and asymmetric drifts of older stars. From F-type to K-type, $\langle V_R \rangle$ increases steadly from $\sim$ 0 to $\sim$ 4~km~s$^{-1}$ for the metal-rich populations, or from $\sim -$6~km~s$^{-1}$ to $\sim$ 2~km~s$^{-1}$ for the metal-poor populations. For each spectral-type, the metal-poor stars always have lower values of $\langle V_R \rangle$ than the metal-rich ones. Similarly, $\langle V_Z \rangle$ decreases towards later spectral-types amongst the metal-rich populations, from $\sim$ 2~km~s$^{-1}$ for the F-type stars to $\sim -$2~km~s$^{-1}$ for the K-type stars. For the metal-poor populations, $\langle V_Z \rangle$ of the G-type stars are lower than that of the F-type by $\sim$ 3~km~s$^{-1}$, but almost equal to that of the K-type. For each spectral-type, metal-poor stars always have larger $\langle V_Z \rangle$ than metal-rich ones.

\begin{table*}
\centering
\begin{minipage}{150mm}
\caption{Bulk motions of the local sample.}
\begin{tabular}{ccrrrrrrr}
\hline
  &  &  &
\multicolumn{3}{c}{UCAC4 Proper Motions} &
\multicolumn{3}{c}{PPMXL Proper Motions} \\
$T_{\rm eff}$ & [Fe/H]  & \multicolumn{1}{c}{Number} & 
\multicolumn{1}{c}{$\langle V_R \rangle$} & \multicolumn{1}{c}{$\langle V_\Phi \rangle -$ 220}  & \multicolumn{1}{c}{$\langle V_Z \rangle$} &
\multicolumn{1}{c}{$\langle V_R \rangle$} & \multicolumn{1}{c}{$\langle V_\Phi \rangle -$ 220}  & \multicolumn{1}{c}{$\langle V_Z \rangle$} \\
(K) & (dex) & \multicolumn{1}{c}{of stars} &
\multicolumn{1}{c}{(km s$^{-1}$)} & \multicolumn{1}{c}{(km s$^{-1}$)} & \multicolumn{1}{c}{(km s$^{-1}$)} &
\multicolumn{1}{c}{(km s$^{-1}$)} & \multicolumn{1}{c}{(km s$^{-1}$)} & \multicolumn{1}{c}{(km s$^{-1}$)} \\
\hline
(6000, 6800) &     $> -$0.4 &    92 &
$-$0.3$\pm$1.1   &  2.2$\pm$0.9        & 1.6$\pm$0.6 &
1.0$\pm$1.3        &  2.2$\pm$0.8        & 2.4$\pm$0.6 \\
(5000, 6000) &     $> -$0.4 & 2473 &
2.2$\pm$0.2         & $-$5.6$\pm$0.2   & $-$0.2$\pm$0.1 &
2.0$\pm$0.2         & $-$5.7$\pm$0.2   & 0.4$\pm$0.1 \\
(4200, 5000) &     $> -$0.4 & 2581 &
4.3$\pm$0.2         & $-$5.6$\pm$0.2   & $-$1.8$\pm$0.1 &
4.1$\pm$0.2         & $-$6.2$\pm$0.2   & $-$1.4$\pm$0.1 \\
(6000, 6800) & $< -$0.4 &    46 &
$-$6.6$\pm$3.0    & $-$0.6$\pm$2.0   & 5.1$\pm$0.5 &
$-$4.8$\pm$3.4    & $-$1.7$\pm$1.9   & 4.5$\pm$0.6 \\
(5000, 6000) & $< -$0.4 &  170 &
$-$3.6$\pm$1.5    & $-$11.9$\pm$1.4  & 1.4$\pm$1.0 &
$-$3.4$\pm$1.6    & $-$14.1$\pm$1.4  & 1.9$\pm$0.9 \\
(4200, 5000) & $< -$0.4 &   81 &
1.3$\pm$2.2         & $-$20.0$\pm$1.9  & 1.8$\pm$1.0 &
1.2$\pm$2.2         & $-$19.3$\pm$2.1  & 1.5$\pm$1.2 \\
\hline
\end{tabular}
\end{minipage}
\end{table*}

\section{Velocity Structures in the Galactic Anti-Center}
\label{sec_popgac}

Stars along the path from the Sun towards the Galactic anti-center have Galactocentric radial velocities $V_R$ aligned with their line-of-sight radial velocities $V_r$. With an analysis of the line-of-sight radial velocities of $\sim$~700 red clump stars in the direction of Galactic anti-center, Liu et al.~(2012) report a ``velocity bifurcation" at Galactocentric radii 10--11~kpc and attribute it to a resonance feature of the central bar. However, the majority stars of their sample fall just around these radii, it is thus possible that the bifurcation may not be a localized feature but rather a global one -- it was only detected at these radii simply because of the limited distance range probed by the sample.

\begin{figure*}
\centering
\includegraphics[scale=0.65]{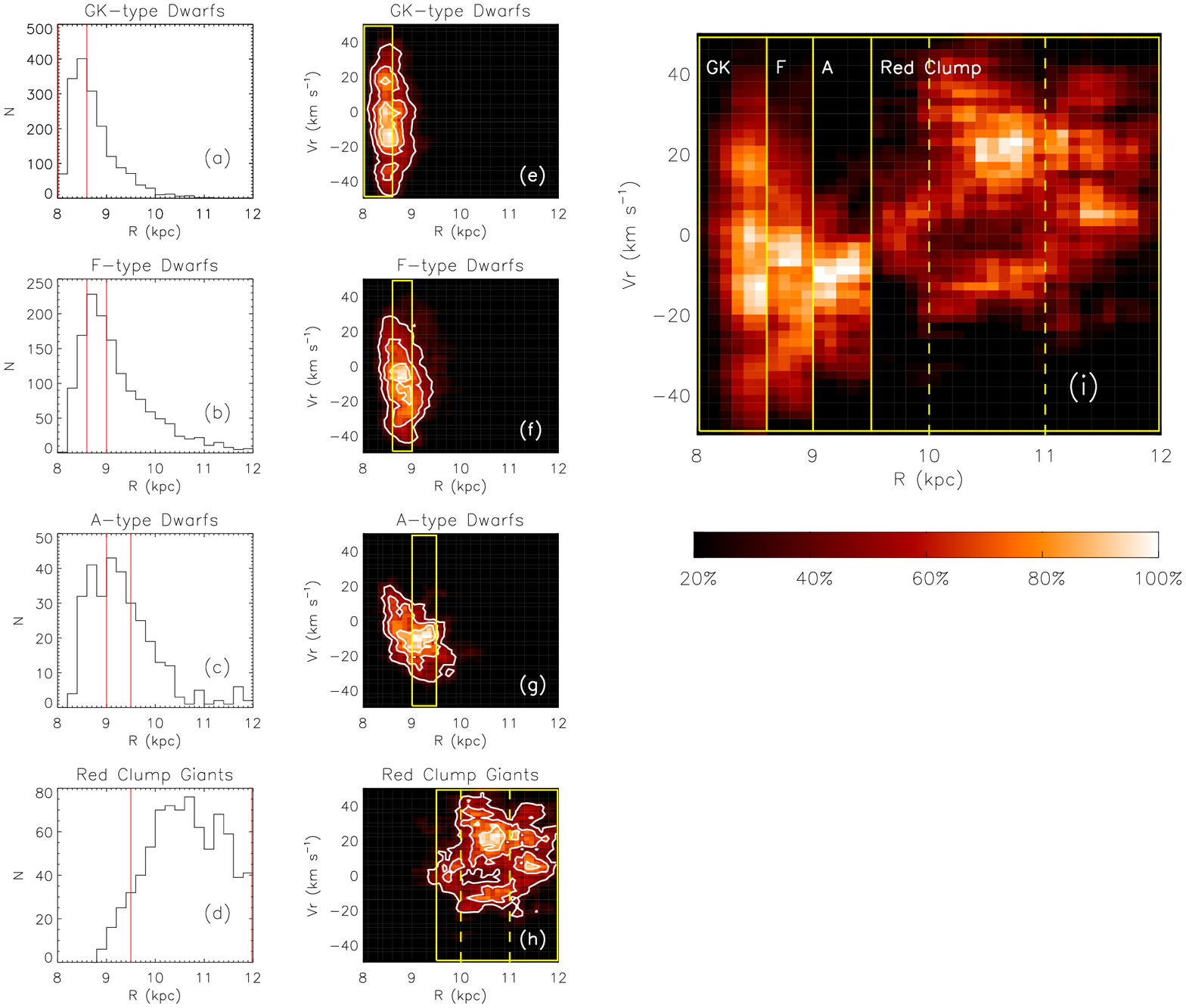}
\caption{Panels (a-d): Distributions of Galactocentric distances of GK-, F- and A-type dwarfs and RC giants in the anti-center sample. The red lines in the four panels mark the distance ranges of the boxed regions in Panels (e-h), respectively. Panels (e-h): Stellar number density distributions of the four populations in the ($R$, $V_r$) plane. The colorscale is proportional to the number of stars in a 500~pc $\times$ 10~km~s$^{-1}$ running box relative to the maximum. The white lines are the corresponding 40\%, 60\% and 80\% contours from outside to inside. Panel (i): Mosaic of the boxed regions in Panels (e-h). The yellow dashed lines and boxes mark the boundaries of distance regimes mentioned in Section~\ref{sec_popgac}.
}
\label{figbif}
\end{figure*}

To clarify the situation, we make use the anti-center sample defined in Section~\ref{sec_acsam}, consisting 1,716 GK-, 1,402 F-, 359 A-type dwarfs and 958 RC giants. Their distance distributions can be found in Panels (a-d) of Fig.~\ref{figbif}. With increasing luminosities, the four groups of stars are able to trace a contiguous distance range from the solar vicinity of $R$~$\sim$~8~kpc out to $R$~$\sim$~12~kpc.

Panels (e-i) show the stellar number density distributions of the anti-center sample in the ($R$, $V_r$) plane. The distributions exhibit a rich variety of velocity structures at different distance regimes. In the first regime, 8.0--8.6~kpc, there are four concentrations of GK-type dwarfs at $V_r$ $\sim$ $-$32, $-$15, 0 and 18~km~s$^{-1}$, respectively. They are apparently related to the known moving groups: the $-$32~km~s$^{-1}$ concentration is linked with Wolf 630, the $-$15~km~s$^{-1}$ one with Sirius, the 0~km~s$^{-1}$ one with Coma Berenices, and the concentration at 18~km~s$^{-1}$ comes from Pleiades (see e.g. Dehnen~1998; Antoja et al.~2012). Moving out to the 8.6--9.0~kpc regime, however, the F-type dwarfs show only two peaks at $V_r$ $\sim$ $-$30 and $-$5~km~s$^{-1}$, respectively. In the third regime, 9.0--9.5~kpc, there is a single concentration of  A-type dwarfs at $V_r$ $\sim$ $-$10~km~s$^{-1}$. Then the concentration moves to $V_r$ $\sim$ 0~km~s$^{-1}$ between 9.5--10~kpc traced by RC stars. In the regime of 10--11~kpc, the ``velocity bifurcation" reported by Liu et al.~(2012) is clearly visible, appearing as two stellar concentrations at $V_r$ $\sim$ 20 and $-$10~km~s$^{-1}$, respectively. The peaking velocities are slightly lower than theirs, by $\sim$~6~km~s$^{-1}$, which possibly originates from the measurement uncertainties. Further out to 11--12~kpc, the RC giants show a new triple-peaked structure, with peaks located at $V_r$ $\sim$ 5~km~s$^{-1}$, 20~km~s$^{-1}$ and 35~km~s$^{-1}$, respectively.

Antoja et al.~(2012) have investigated the moving groups in different regions of the disk, showing that they drift in the velocity space at different radii. Similarly, the anti-center sample reveals a contiguous spatial variations of velocity structures, from the solar vicinity out to $R$ = 12~kpc. Specially, the anti-center sample analyzed here shows that while the ``velocity bifurcation" reported by Liu et al.~(2012) is indeed a localized feature, it is also not the sole feature in the direction of Galactic anti-center. It is thus not an obvious conclusion that the ``velocity bifurcation" is a resonance feature of the central bar, and the formation mechanisms of these features need further investigations.

\section{Summary}
\label{sec_sum}

In this paper we have carried out a detailed investigation of bulk motions of the nearby Galactic stellar disk using three samples selected from the LSS-GAC DR2: a global sample of $\sim$~0.57 million FGK dwarfs out to a distance of $\sim$~2~kpc, a local sample of $\sim$~5,400 FGK dwarfs with distances $\leq$~150~pc, and an anti-center sample containing $\sim$~4,400 AFGK dwarfs and RC giants extending out to $\sim$~4~kpc in the Galactic anti-center direction.

With the global sample, we have provided a three-dimensional map of bulk motions of the Galactic disk from the solar vicinity out to $\sim$~2~kpc with a spatial resolution of $\sim$~250~pc. Typical values of radial and vertical bulk motions range from $-$15 to 15~km~s$^{-1}$, while the lag behind the circular speed dominates the azimuthal bulk motion by up to $\sim$~15~km~s$^{-1}$. The map reveals spatially coherent, kpc-scale stellar flows, with typical velocities of a few tens~km~s$^{-1}$. We have analyzed the bending and breathing modes and higher-order perturbations. The bending and breathing modes are clearly visible, and vary smoothly across the disk plane. The data also reveal breaks and ripples in the profiles of vertical bulk motion versus height, indicative of higher-order perturbations.

With the local sample, we have studied the bulk motions of different stellar populations. Differences of several km~s$^{-1}$ are found in the radial and vertical bulk motions, while the azimuthal bulk motions exhibit larger differences.

The anti-center sample reveals abundant velocity structures at distances ranging from the solar vicinity out to a Galactocentric radius of 12~kpc, with the nearer ones apparently related to the known moving groups. The ``velocity bifurcation" reported by Liu et al. is confirmed to be a localized feature at 10--11~kpc, beyond which the data reveal a new triple-peaked structure.

\begin{acknowledgements}
This work is supported by National Key Basic Research Program of China 2014CB845700. This work has used data products from the Guoshoujing Telescope (the Large Sky Area Multi-Object Fiber Spectroscopic Telescope, LAMOST). LAMOST is a National Major Scientific Project built by the Chinese Academy of Sciences. Funding for the project has been provided by the National Development and Reform Commission. LAMOST is operated and managed by the National Astronomical Observatories, Chinese Academy of Sciences.
\end{acknowledgements}

\label{lastpage}

\end{document}